\documentclass[%
 reprint,
superscriptaddress,
twocolumn,
 amsmath,amssymb,
 aps,
prl,
]{revtex4-2}

\usepackage{booktabs}
\usepackage{graphicx}
\usepackage{dcolumn}
\usepackage{bm}
\usepackage{xcolor}
\usepackage{float}
\usepackage{microtype}
\usepackage{bbm}
\usepackage{amsmath}
\usepackage{amsfonts}
\usepackage{braket}
\usepackage{txfonts}
\usepackage[normalem]{ulem}

\usepackage{hyperref}
\hypersetup{colorlinks=true,breaklinks,linkcolor=blue,urlcolor=blue,citecolor=blue}
\usepackage{orcidlink}

\newcommand\AddAuthorComment[3]{
    {\color{#1} ({\bf #2}%
        \if\relax\detokenize{#3}\relax%
        \else%
            {\normalfont: #3}%
        \fi%
    )}%
}
\newcommand\AuthorReplace[5]{
    \AddAuthorComment{#1}{#2}{#3}\linebreak[1]%
    \if\relax\detokenize{#4#5}\relax%
    \else%
        { \color{#1}\sout{#4}\linebreak[1]\uwave{#5}}%
    \fi%
}

\definecolor{darkgreen}{rgb}{0,0.5,0}
\newcommand{\changed}[1]{{\color{darkgreen}{#1}}} 

\newcommand\Dmax{\ensuremath{D_\mathrm{max}}}
\newcommand\Dsum{\ensuremath{D_\mathrm{sum}}}

\renewcommand\section[1]{\paragraph{#1.}}

\begin{document}

\preprint{APS/123-QED}

\title{
Diagnosing phase transitions through time-scale entanglement
}

\author{Stefan Rohshap\,\orcidlink{0009-0007-2953-8831}}
\email{stefan.rohshap@tuwien.ac.at}
\affiliation{%
    Institute of Solid State Physics, TU Wien, 1040 Vienna, Austria}%

\author{Hirone Ishida\,\orcidlink{0009-0000-7095-7105}}
\affiliation{%
    Department of Physics, Saitama University, Saitama 338-8570, Japan}%

\author{Frederic Bippus\,\orcidlink{0009-0006-4316-6547}}
\affiliation{%
    Institute of Solid State Physics, TU Wien, 1040 Vienna, Austria}%

\author{Leonard M. Verhoff\,\orcidlink{0009-0004-3358-1312}}
\affiliation{%
    Institute of Solid State Physics, TU Wien, 1040 Vienna, Austria}%

\author{Anna Kauch\,\orcidlink{0000-0002-7669-0090}}
\affiliation{%
    Institute of Solid State Physics, TU Wien, 1040 Vienna, Austria}%
    
\author{Karsten Held\,\orcidlink{0000-0001-5984-8549}}
\affiliation{%
    Institute of Solid State Physics, TU Wien, 1040 Vienna, Austria}

\author{Hiroshi Shinaoka\,\orcidlink{0000-0002-7058-8765}}
\affiliation{%
    Department of Physics, Saitama University, Saitama 338-8570, Japan}%

\author{Markus Wallerberger\,\orcidlink{0000-0002-9992-1541}}
\affiliation{%
    Institute of Solid State Physics, TU Wien, 1040 Vienna, Austria}%

\date{\today}

\begin{abstract}
Spatial entanglement of quantum states has become a central paradigm of many-body physics. Here, we unearth a fundamentally different form of entanglement, the entanglement between imaginary time scales. This time-scale entanglement is accessible through quantics tensor train diagnostics (QTTD), where the bond dimension of an $n$-particle correlator encodes the coupling between temporal scales. Our central result is that time-scale entanglement is generically enhanced in the vicinity of phase transitions and crossovers. At quantum critical points, it becomes scale-invariant.
We demonstrate time-scale entanglement  across a range of systems, including finite-size Hubbard rings, the transverse-field Ising model, the single-impurity Anderson model, 
and the Mott transition in the Hubbard model.
Remarkably, the enhanced time-scale entanglement is largely independent of the specific observable, establishing QTTD as a universal and unbiased diagnostic of criticality. 
\end{abstract}

\maketitle

\section{\label{sec:introduction}Introduction}

Phase transitions are among the most intriguing phenomena in physics. Yet, their numerical treatment and detection can be challenging, especially when exotic phases emerge \cite{Landau1937,Geffroy2018,Niyazi2020}. In extended many-body systems, the full wave function is usually not available, and using susceptibilities to diagnose transitions can be brittle, since we need to know {\em a priori} where to look, i.e., we must make sure to compute the response associated with the transition in question or at least one that is not insensitive to the transition.
In general, we can expect some but a much weaker signal in the susceptibilities that are not the one associated with the symmetry breaking for a second-order phase transition. 
On the other hand, susceptibilities generally provide less clear precursors near first-order phase transitions.

Similarly, entanglement is one of the most enigmatic quantum phenomena. It is challenging to calculate in true many-electron systems where many electrons contribute and are entangled~\cite{Amico_2008,Laurell_2025,Horodecki_2009,balut2025quantumfisherinformationreveals,mazza_quantum_2024,bellomia_quasilocal_2024,walsh_entanglement_2020}.  In (quasi-) one-dimensional spin systems, matrix product states (MPS)~\cite{Affleck87,Schollwoeck11} have unlocked much of the theoretical and numerical analysis of entanglement~\cite{Cirac_2021,Eisert_2010} and its growth~\cite{Hastings_2007,He_2017,Wolf_2014}. MPS add to the physical space an auxiliary, {\em latent} space, whose dimension (``bond dimension'') indicates the strength of spatial entanglement~\cite{Verstraete_2006, Li2024}.

Recently, it was shown that the machinery of MPS can be reused to compactify arbitrary functions in space and time~\cite{Oseledets2009, Oseledets2011, Khoromskij2011, Dolgov2012, Khoromskij2018}. Instead of a physical spin at some site, each tensor represents a different time/length scale of the function, so as we step along the train, we zoom in and out rather than move left or right. This ansatz was termed quantics tensor train (QTT). It was instrumental in breaking computational barriers in modeling turbulence~\cite{Gourianov2022-vn,peddinti2023complete, kornev2023,holscher2024, Gourianov2024}, plasmas~\cite{Ye2022}, quantum chemistry~\cite{Jolly2023}, and -- most relevant to this Letter -- electronic correlation functions.  In this latter case, some of us showed how to sample~\cite{Ritter2024}, store~\cite{Shinaoka2023}, and perform calculations~\cite{Rohshap2024} with electronic response functions which were previously inaccessible. QTTs by construction also feature a latent space and an associated bond dimension, which has hitherto been chiefly used as a measure of performance.

\begin{figure}
    \setlength{\abovecaptionskip}{0pt}
    \setlength{\belowcaptionskip}{-13.5pt}
    \centering 
    \includegraphics[width=\linewidth]{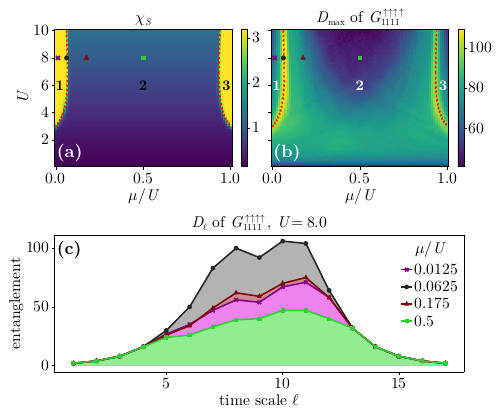}
    \caption{Entanglement of exponentially different imaginary time scales for the Hubbard dimer at $T=0.02$: {\bf (a)} Local spin susceptibility $\chi_s$ indicating different ground states filled with one, two and three electrons; {\bf (b)} QTT maximum bond dimension $D_{\max}:= \max_{\ell} D_{\ell}$ of the imaginary time four-point Green's function $G_{1111}^{\uparrow \uparrow \uparrow \uparrow}$ as a function of interaction strength $U$ and chemical potential $\mu/U$, where red-dashed lines mark crossings of the ground state; and {\bf (c)} QTT bond dimension $D_\ell$ between timescale $2^{-\ell}$ and $2^{-\ell -1}$ 
    for different values of $\mu/U$ at $U=8$, marked by the same symbol and color in (a) and (b).
    }
    \label{fig:Hubbard-dimer}
\end{figure}

In this Letter, we reason by analogy with MPS, where the bond dimension gauges spatial entanglement: we propose the QTT bond dimension of electronic functions as a physical measure, namely as the strength of time and length scale entanglement. For the example of the Hubbard dimer, we plot this over different time scales in Fig.~\ref{fig:Hubbard-dimer}(c).

Here, we observe that time-scale entanglement is maximal at crossovers and phase transitions [red dots in Fig.~\ref{fig:Hubbard-dimer}(b)]. These maxima can appear in both single- 
and two-particle propagators [Fig.~\ref{fig:Hubbard-dimer} (b)], without the need to identify the specific susceptibility associated with the transition.  We call this program of detecting phase transitions {\em quantics tensor train diagnostics} (QTTD). 
Additional results and details supporting the conclusions reached for all models discussed are provided in the Supplemental Material (SM)~\cite{SupplementalMaterial}.

\section{Quantics tensor trains}
In QTTs~\cite{Oseledets2009,Shinaoka2023} each variable is represented through a set of binary  numbers or ``quantics'' corresponding to different length or time scales. The resulting tensor is then factorized into a tensor train (TT) at each scale. For illustration, consider 
a function $f(\tau)$ of a discretized variable $\tau$, with $\tau \in \{0, \ldots, M-1\}$
on $M = 2^R$ grid points. In  quantics, $\tau$ is expressed in  binary representation $\tau = (\sigma_1 \sigma_2 \dots \sigma_R)_2 = \sum_{\ell=1}^{R} 2^{R-\ell} \sigma_{\ell},$ with $\sigma_\ell \in \{0, 1\} \, $ leading to $f(\tau)$ being seen as a $2\times 2\times ...\times 2$ ($R$ times) tensor \(F_{\sigma_1, \ldots, \sigma_R}\) instead. Now each tensor index $\sigma_\ell$ corresponds to an exponentially distinct length or time scale of the system. The first bit \(\sigma_1\) represents the coarsest scale which divides the system into halves, while the last bit $\sigma_R$ reflects the finest scale.

Factorization of this tensor into a (truncated) tensor train (TT) or matrix product state (MPS) of the form
\vspace{-2mm}
\begin{equation}
    F_{\sigma_1 \ldots \sigma_R} \approx \tilde F_{\sigma_1 \ldots \sigma_R} =
    \!\sum_{\alpha_1=1}^{D_1}\!\ldots\!\sum_{\alpha_{R-1}=1}^{D_{R-1}} [M_1]^{\sigma_1}_{1\alpha_1} [M_2]^{\sigma_2}_{\alpha_1\alpha_2} \cdots [M_R]^{\sigma_R}_{\alpha_{R-1} 1}
    \label{eq:TTdecomposition}
\end{equation}
can be achieved with singular value decomposition (SVD) \cite{Shinaoka2023} or using  tensor cross interpolation (TCI)  \cite{NunezFernandez2022,Ritter2024,NunezFernandez2024,Ritter2026}. In Eq.~\eqref{eq:TTdecomposition}, each \(M_\ell\) is a $D_{\ell-1} \times 2 \times D_\ell$ tensor with ``physical'' binary index \(\sigma_\ell\) and virtual indices (``bonds'') \(\alpha_{\ell-1}, \alpha_{\ell}\), which are summed over.  Eq.~\eqref{eq:TTdecomposition} translates to:
\begin{equation}
    \raisebox{-2em}{%
        \includegraphics[scale=0.9]{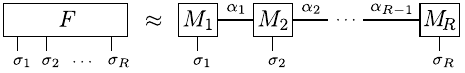}
    }
    \label{eq:TTdiagram}
\end{equation}
in tensor network form~\cite{Schollwoeck11}.
We define the (maximum) bond dimension of the QTT as $\Dmax=\textrm{max}_\ell(D_\ell)$. Generally,  bond dimensions \(D_\ell\) characterize the flow of information between different time scales (as well as length scales) and, thus, the amount of ``length or time-scale entanglement'' in the system 
up to some precision $\epsilon$~\cite{Rohshap2025}. In case of SVD, usually the squared Frobenius norm $\epsilon = ||F- \tilde F||^2_{\mathrm{F}} / ||F||^2_{\mathrm{F}}$ is taken as a measure of the error of the approximation in Eq.~\eqref{eq:TTdecomposition}. 
Many problems in many-body physics have been shown to use quantities that are strongly compressible into QTTs with small bond dimension
~\cite{Shinaoka2023, Ishida2024, Murray2024, takahashi2024compactnessquanticstensortrain, Ritter2024, Rohshap2024,Erpenbeck2023, Eckstein2024, Sroda2024, Frankenbach2025, Rohshap2025,Grosso2026}. 

\section{Quantics tensor trains diagnostics}

Let us start from a general imaginary time two-point correlator $G_{AB}(\tau)=-\langle \hat{\mathcal{T}} \hat A(\tau) \hat B \rangle$ with fermionic operators $\hat A, \hat B$ and the time ordering operator $\hat{\mathcal{T}}$. For a finite or gapped system, $G_{AB}(\tau)$ can be expressed in its Lehmann representation:
\begin{equation}
    G_{AB}(\tau) = - \frac{1}{Z} \sum_n e^{-\beta \epsilon_n} \sum_m e^{-\tau(\epsilon_m - \epsilon_n)} \langle n | \hat A | m \rangle \langle m | \hat B  | n \rangle,
    \label{eq:lehmann}
\end{equation}
where $1/\beta$ is temperature, $\tau \in [0,\beta)$ is imaginary time, $Z$ is the partition function, $\hat H$ is the Hamiltonian including the chemical potential, and $\hat H|n\rangle = \epsilon_n|n\rangle$ defines its eigenenergies and eigenstates. 
Clearly, $G_{AB}(\tau)$ is the sum of exponentials. An exponential has a QTT representation of bond dimension one, since it is a product of exponentials of the binary variables $\sigma_\ell$ at each scale: $e^{\tau \epsilon} = \prod_{\ell=1}^{R} e^{\epsilon\beta 2^{-\ell} \sigma_{\ell}}$. The bond dimension of the exact QTT representation of $G_{AB}$ is, thus, bound by the number of nonzero elements in the sum. Let us now consider the case of temperature approaching zero ($\beta \rightarrow \infty$), then
\vspace{-0.25cm}
\begin{align} \label{eq:two-point-correlator-Lehmann-zero-T-crossing}
    G_{AB}(\tau) &= -  \frac{1}{N} \sum_{i=1}^N \sum_m \Big( e^{-\tau \epsilon_m} \langle \mathrm{GS}_i | \hat A | m \rangle \langle m | \hat B  | \mathrm{GS}_i \rangle \nonumber  \\
    &\quad + e^{-(\beta-\tau) \epsilon_m} \langle m | \hat A | \mathrm{GS}_i \rangle \langle \mathrm{GS}_i | \hat B  | m \rangle \Big), 
\end{align}
where $|\mathrm{GS}_i \rangle$ denotes the $i-$th ground state and $N$ is the number of ground states (states with the lowest energy). We set $\epsilon_{\mathrm{GS}}=0$; and the first (second) term contributes predominately for small (large) values of $\tau$.

For a single non-degenerate ground state,
the bond dimension of the exact QTT representation is bounded by the number of nonzero terms in the sum, maximally by $M-1$  for $M$ eigenstates $m$. Next, consider a ground state crossing of the two lowest states $|\mathrm{GS}_1\rangle$ and $|\mathrm{GS}_2 \rangle$ ($N=2$). Then the bond dimension of the exact QTT is at most $2M-2$ since twice as many exponentials can contribute. In practice, it may be lower as fewer states $| m \rangle$ will give nonzero matrix elements. Still, at the ground state crossing the sum consists of the exponentials and matrix elements associated with $|\mathrm{GS}_1 \rangle$ and $|\mathrm{GS}_2 \rangle$ leading to larger bond dimensions of $G_{AB}(\tau)$ than before or after the crossing, as long as $\hat A$ and $\hat B$ are not completely orthogonal to the response of the system.
Therefore, at zero temperature, a ground state crossing in finite systems
can be determined by monitoring the bond dimension of the QTT representation. We call this approach {\it quantics tensor train diagnostics} (QTTD), where the diagnostics power of the approach is further examined in Apps. A,B. 

Consider finite, but low temperatures. Then, also excited states $|n \rangle$ contribute to $G_{AB}(\tau)$, suppressed by $e^{-\beta \epsilon_n}$. 
Most of these contributions will be truncated, if they are below a certain specified cutoff in the QTT construction 
of the two-point correlator. Hence, we will still see a peak of 
\Dmax{} in the vicinity of the corresponding crossing (at $T=0$), 
that will be smeared out with increasing temperature. 

We can summarize the QTTD procedure in the following way. i) Calculate an available correlator of the system. ii) Vary the QTT cutoffs and analyze the bond dimension profile to find stable features corresponding to possible phase transitions or crossovers (see App. C for more details). iii) Iterate this procedure for different available correlators to verify the obtained maxima as universal features corresponding to the system-inherent rise in time-scale entanglement associated with a phase transition or crossover. 

In the following, we use the notion of imaginary time and Matsubara frequency scale entanglement interchangeably, as both representations are related by a Fourier transform and encode the same underlying multi-scale structure of the correlation functions. Before investigating genuine phase transitions in extended systems, let us first verify the validity of this derivation in simple finite systems.

\section{Hubbard model}

We start by introducing the Hubbard model on $N$ sites:
\begin{equation}
    \hat H = -\sum_\sigma \sum_{i,j=1}^N t_{i-j} \hat c_{i,\sigma}^{\dagger} \hat c_{j,\sigma} 
    + \sum_{i=1}^N \left( U  \hat n_{i, \uparrow} \hat n_{i,\downarrow} - \mu (\hat n_{i,\uparrow} + \hat n_{i,\downarrow}) \right),
     \label{eq:general-Hubbard-Hamiltonian}
\end{equation}
where $t_{\pm1} \equiv t$ and $t_{\pm2} \equiv t'$ are the nearest and next-nearest hopping amplitude, respectively, all other hopping amplitudes are set to zero,
the local on-site Coulomb interaction is $U$, the chemical potential is $\mu$, and the number operators $\hat n_{i, \sigma} = \hat c_{i,\sigma}^{\dagger} \hat c_{i,\sigma}$ are defined via the fermionic annihilation (creation) operators $\hat c_{i,\sigma}^{(\dagger)}$ with $\sigma=\uparrow,\downarrow$ and site index $i$. We use $t \equiv k_B \equiv 1$ to set energy and temperature units. 

\section{Hubbard dimer}
For the Hubbard dimer [\changed{1D,} $N=2, t' = 0$ in Eq.~\eqref{eq:general-Hubbard-Hamiltonian}], all 16 eigenstates can be analytically calculated~\cite{Wallerberger2022}, providing a first good test case for QTTD. 
Fig.~\ref{fig:Hubbard-dimer}(b) shows \Dmax{} of the local two-particle Green's function $G_{1111}^{\uparrow \uparrow \uparrow \uparrow}(\tau_1, \tau_2, \tau_3) = -\langle \mathcal{T} \hat c_{1,\uparrow} (\tau_1) \hat c_{1,\uparrow}^{\dagger} (\tau_2) \hat c_{1,\uparrow} (\tau_3) \hat c_{1,\uparrow}^{\dagger} \rangle$ with $\epsilon = 10^{-14}$ at $\beta=50$ and $R=6$. As expected a sharp peak in \Dmax{} can be seen in the vicinity of the ground state crossing, indicated by dashed lines. Here, the singlet ground state crosses with a doublet that has one (three) electrons at small (large) values of $\mu/U$. In Fig.~\ref{fig:Hubbard-dimer}(a), the value of the spin susceptibility $\chi_S$ (definition in App. D) is shown, clearly distinguishing the two ``phases''. Fig.~\ref{fig:Hubbard-dimer}(c) plots the bond dimension as time-scale entanglement measure over the imaginary time scale.
Additionally, we extend the analysis to four-site Hubbard rings with and without nearest neighbor hoppings~\cite{Nishimoto2008} in the SM~\cite{SupplementalMaterial}. Besides varying QTT cutoffs to identify ground state crossings and thermal crossovers, various entanglement measures~\cite{Wang2015,Zanardi2006, You2007, Zanardi2007, Venuti2007, Kashihara2023, Wang2015b,hauke_measuring_2016,Qu_phase_transition_FQ,zurek_information_1983,barnett_entropy_1989,roosz_two-site_2024, Bippus2025a, Bippus2025b,grover_entanglement_2013,renyi_measures_1961,peres_separability_1996,horodecki_separability_1996,DeChiara_2018,ZHENG_negativity_phase_transition,Das_I_phase_transition,Yamashika2025,Bellomia_entanglment_MI,bellomia2025localclassicalcorrelationsphysical} are analyzed in the SM~\cite{SupplementalMaterial}.

\section{Quantum criticality in transverse-field Ising model}
\begin{figure}
\setlength{\abovecaptionskip}{0pt}
    \setlength{\belowcaptionskip}{-14.5pt}
    \raggedleft
\includegraphics[width=1.0\linewidth]{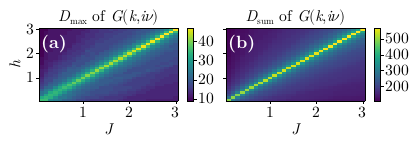}
        \par\vspace{-2.5mm}
\includegraphics[width=1.0\linewidth]{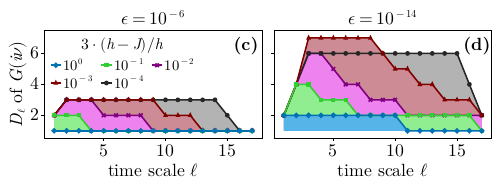}
\caption{\Dmax{} (a) and \Dsum{} (b) of $G(k,i\nu)$ of the TFIM for $R=12,\epsilon=10^{-14}$ at $\beta=10^4$. (c), (d): Bond dimension profile $D_\ell$ of local Green's function $G(i \nu)$ at $\beta=10^7$ near criticality ($h=3$). Criticality and scale invariance lead to uniform entanglement across logarithmic time scales independent of $\epsilon$.}
    \label{fig:TFIM}
\end{figure}

At a quantum phase transition (QPT), the system becomes scale invariant in both spatial and imaginary-time directions. In the absence of a characteristic scale, fluctuations span all temporal scales, resulting in comparable contributions from coarse and fine scales to the system’s response. In the following, we investigate how this emergent scale invariance manifests itself in the structure of time-scale entanglement in the vicinity of the QPT of the one-dimensional transverse-field Ising model (TFIM), described by $\hat H = -\sum_{i=1}^L(J \hat \sigma_i^x \hat \sigma_{i+1}^x +h \hat \sigma_i^z)$ with the spin operators $\hat \sigma_i^{\alpha}$,  nearest-neighbor spin-coupling $J$ and a transverse field $h$. At zero temperature, the model exhibits a continuous quantum phase transition between ferromagnetic and paramagnetic phases at $J=h$, where the system becomes scale invariant in both spatial and imaginary time directions. Exploiting the exact analytical solution for the normal Green’s function $G(k,i\nu)$ (see App.~F), we perform a QTT compression at very low temperature ($\beta=10^4$). As shown in Fig.~\ref{fig:TFIM}(a,b), both the maximum bond dimension \Dmax{} and, more prominently, the sum of bond dimensions, $D_{\mathrm{sum}}:= \sum_{\ell} D_{\ell}$, which quantifies the total coupling between scales, exhibit clear peaks at the critical point $J=h$. This reflects the increasing complexity of the correlation structure as the system approaches criticality, where contributions from all scales become relevant.
A physically interesting signature of criticality emerges in the bond dimension profile of the local Green’s function $G(i\nu)$ at ultra-low temperatures ($\beta=10^7$) very close to the QPT. As shown in Fig.~\ref{fig:TFIM}(c,d), the bond dimensions become nearly independent of the time scale $\ell$, indicating that all imaginary time scales contribute equally when approaching criticality. This flattening of the bond dimension profile constitutes a direct numerical manifestation of scale invariance: the absence of a characteristic temporal scale at the QPT is encoded as a uniform distribution of entanglement across logarithmic time scales. Notably, this behavior is robust with respect to the QTT compression threshold $\epsilon$, affecting only the overall magnitude of the bond dimensions but not their scale-independent structure. These findings are closely connected to earlier results for critical quantum systems, where entanglement grows logarithmically with system size, reflecting an organization of correlations across length scales that is naturally captured by the multiscale entanglement renormalization ansatz (MERA)~\cite{Vidal2003, Vidal2007, Vidal2008, Crosswhite2008,Pfeifer2009,Pirvu2012,Haegeman2013,Vanhecke2019,Acoleyen2020}. In this context, our results suggest an analogous multi-scale organization in imaginary time, directly revealed through QTT bond dimensions.

%

\section{Mott transition}
\begin{figure}
    \setlength{\abovecaptionskip}{0pt}
    \setlength{\belowcaptionskip}{-12.5pt}
    \centering 
    \includegraphics[width=\linewidth]{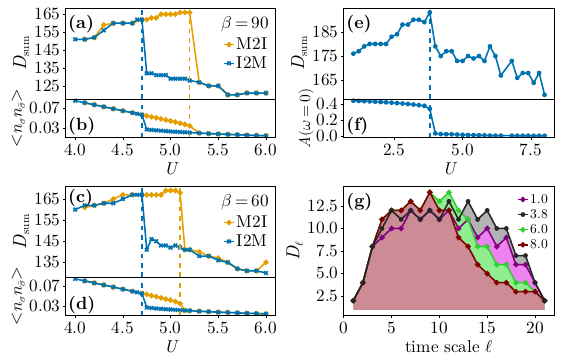}
    \caption{Sum of QTT bond dimensions of $G(i\nu)G(i\nu+i\omega)$ obtained from DMFT solution for the Bethe lattice for $\beta=90$ (a) and $60$ (c) initialized in insulating (I2M) and metallic (M2I) states vs. the double occupancy $\langle n_\sigma n_{\bar \sigma} \rangle$  ((b), (d)) for various $U$ in units of $t=1$. (e) \Dsum{} of same property for 3D $\textrm{NdNiO}_2$ at $\beta=38$eV$^{-1}$ in comparison to analytically continued spectral weight $A(\omega)$ at $\omega=0$ in (f). (g): Bond dimension profile for various $U$, with spread of time-scale entanglement to fine scales at crossover ($U=3.8$\,eV). QTT parameters: $R=11,\epsilon=10^{-4}$.}
    \label{fig:Mott-QTTD}
\end{figure}

Next, we investigate whether time-scale entanglement, as captured by QTT decompositions, can diagnose phase transitions in realistic many-body calculations. To this end, we analyze the first-order Mott metal-to-insulator transition present in the dynamical mean-field theory (DMFT)~\cite{Metzner1989,Georges1992,Georges1996,w2dynamics} solution of the Hubbard model on the Bethe lattice. Using data from Refs.~\cite{Watzenboeck2022a,Watzenboeck2022}, we construct the particle–hole bubble $G(i\nu)G(i\nu+i\omega)$~\footnote{Enhancement of time-scale entanglement can also be observed in the plain one-particle Green's function $G(i \nu)$, but becomes more explicit in the particle-hole bubble through enhanced scale mixing.} and compress it into a QTT via SVD. Figs.~\ref{fig:Mott-QTTD}(a) and (c) show the resulting \Dsum{} obtained from DMFT solutions initialized in the insulating (I2M) and metallic (M2I) states at $\beta=90$ and 60, respectively, as a function of the interaction strength $U$. In both cases, pronounced peaks in \Dsum{} appear at different values of $U$, coinciding with steep drops in the double occupancy (panels (b) and (d)) that mark the onset of the insulating regime. In contrast to a QPT, a finite temperature first-order transition is characterized by the coexistence of distinct phases with finite correlation lengths and, hence, well-defined characteristic scales. No scale invariance emerges in this case. However, time-scale entanglement is enhanced when approaching the phase transition, both from the metallic and insulating regimes. For a more concise picture, we refer the reader to Fig.~19 in the SM~\cite{SupplementalMaterial} showing an extended $U$-range.

We next extend the analysis to a realistic material computation. Fig.~\ref{fig:Mott-QTTD}(e) shows \Dsum{} of the same bubble quantity computed for 3D $\textrm{NdNiO}_2$ ~\cite{Verhoff2025} within (single-orbital) DMFT calcualtion with density functional theory input~\cite{wien2k,wannier90,wien2wannier,w2dynamics} at $\beta=38 \mathrm{eV}^{-1}$ and half-filling, as a function of $U$, alongside the analytically continued~\cite{ana_cont} spectral weight $A(\omega)$ at $\omega=0$ in panel (f). In contrast to the Bethe lattice at lower T (with respect to bandwidth), no hysteresis is observed. Instead, \Dsum{} exhibits a single pronounced peak that coincides with the suppression of $A(\omega=0)$, signaling a crossover from metallic to insulating behavior. At this crossover ($U\approx 3.8$), the bond dimension profile (panel (g)) reveals a more uniform distribution across scales, indicating enhanced entanglement between imaginary time scales~\footnote{Zigzag of bond dimensions caused by different inter- and intra-scale entanglement between variables in interleaved QTT representation.}. We interpret this observation in the following way. In crossover regimes, no singular behavior or diverging correlation length is present. Instead, spectral weight and correlations are continuously redistributed across energy and time scales leading to broader and less pronounced increases in entanglement of temporal scales. However, near underlying critical points this redistribution can involve a broad range of scales. Especially in the vicinity of a finite-temperature second-order phase transition the emergence of scale invariance in the spatial degrees of freedom may induce broad multi-scale structures in imaginary time, which are encoded in the corresponding correlation functions. As a result, temporal correlations become distributed across a wider range of scales, leading to a partial flattening of the bond dimension profile reflecting the increase of entanglement between temporal scales. Therefore, QTTD suggests a route to probing how scale invariance is encoded in generic observables through enhanced coupling between imaginary time scales. The spectral function is shown in Fig.~\ref{fig:spectral_NNO} in App.~E.

\section{Single-impurity Anderson model}

The single-impurity Anderson model (SIAM) features a magnetic impurity in a metallic bath and shows a crossover between local moment and Kondo regime. For numerical calculations, the continuous bath is discretized leading to the following Hamiltonian
\vspace{-1.5mm}
\begin{small}
    \begin{align}
    \hat{H} =& (\varepsilon_0 - \mu)(\hat n_\uparrow + \hat 
 n_\downarrow) + U \hat n_\uparrow \hat  n_\downarrow +\sum_{\ell=1}^{3} E_{\ell} \hat  c_{\ell}^\dag \hat  c_{\ell} + \sum_{\ell=1}^{3} (V_{\ell} \hat  d_\sigma^\dag \hat  c_{\ell} + \mathrm{h.c.}).
\end{align}
\end{small}
\vspace{-4.5mm}

\begin{figure}
    \setlength{\abovecaptionskip}{0pt}
    \setlength{\belowcaptionskip}{-14.5pt}
    \centering 
    \includegraphics[width=\linewidth]{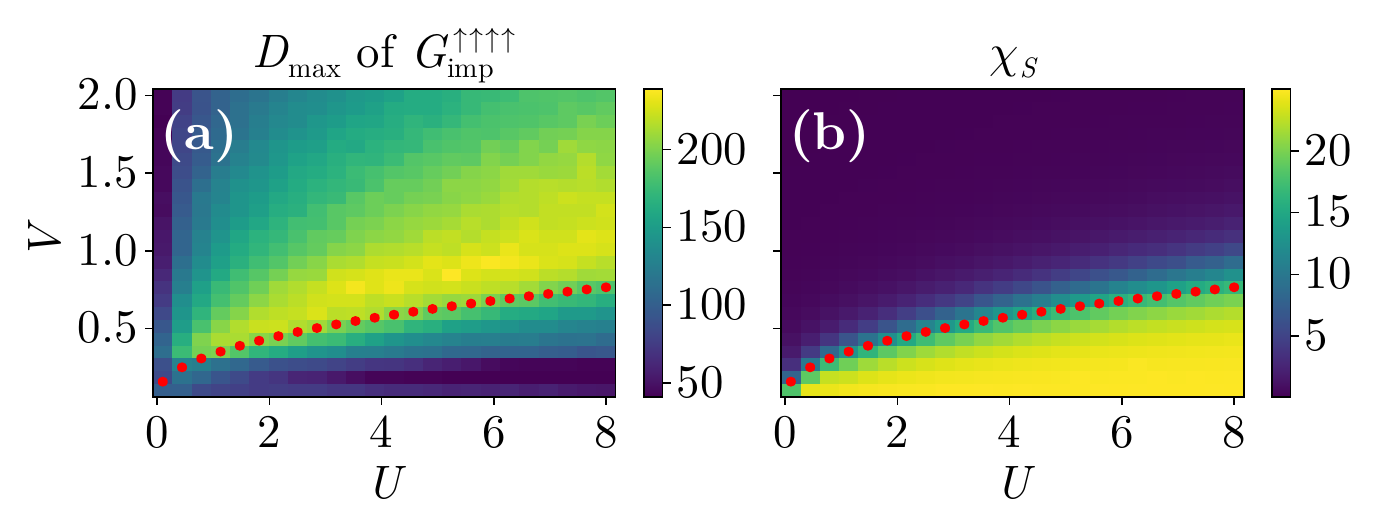}
    \caption{(a)  \Dmax{} of two-particle Green's function of SIAM ($\beta=100$) in $U$-$V$ plane [red dots indicate $T_K(U,V)=1/\beta$]. (b) Spin susceptibility $\chi_{S}$ indicating different regimes.}
    \label{fig:SIAM-QTTD}
\end{figure}

\noindent $\varepsilon_0$ represents the impurity site energy, $\mu$ is the chemical potential, and $U$  the on-site interaction. The annihilation (creation) operator on the impurity site is  $\hat  d_\sigma^{(\dag)}$, and  $\hat  n_{\sigma} = \hat  d_{\sigma}^\dag \hat 
 d_{\sigma}$. The bath density of states is assumed to be semicircular, and the parameters $V_\ell$ and $E_\ell$ denote the hybridization strength and energy levels of the discretized bath (three levels), respectively. 
Figure~\ref{fig:SIAM-QTTD}(a) illustrates \Dmax{} of the QTT representation of  $G^{\uparrow\uparrow\uparrow\uparrow}_{\mathrm{imp}}(\tau_1, \tau_2, \tau_3)$ at the impurity site for $\beta=100$, plotted against the model parameters $U$ and $V$, where $V$ denotes the hybridization strength in the original continuous model. In Fig.~\ref{fig:SIAM-QTTD}(b), the spin susceptibility $\chi_S$ is depicted. The red dots mark the Kondo temperature $T_K=V\sqrt{U}e^{-\pi U/8V^2}$ estimated via poor man’s scaling~\cite{haldane1978scaling, Wang2015b}, which delineates the boundary between  Kondo and local moment regimes, as also reflected in the spin susceptibility behavior. \Dmax{} has a broad peak located in the vicinity of the Kondo temperature, diagnosing the thermally driven Kondo to local moments regime crossover.

\section{Conclusions}

Since their recent 
invention, QTTs have primarily been applied as a compression tool in numerical calculations to overcome severe memory bottlenecks. In this work, we present a novel QTTD approach, diagnosing imaginary time-scale entanglement from the QTT bond dimension. By studying various models we show that time-scale entanglement becomes maximal at phase transitions and crossovers making their identification possible via QTTD.
This is even true for the one-particle Green's function whose static, time-independent part does not signal a second-order phase transition when approaching it from the symmetry-unbroken side. In Green's function dynamics, the phase transition is, however, encoded in its time-scale entanglement. Specifically, we identify distinct signatures of scale invariance and complexity imprinting themselves on generic correlation functions at critical points becoming accessible through QTTD. This implies that time-scale entanglement is a system-inherent property. We hope that this will lead to a better understanding of how criticality and scale invariance manifest in different observables of the system.

Apart from a fresh perspective on phase transitions, and possibly other physical phenomena, the strong point of QTTD is that it can be used universally for zero and finite temperature and multiple correlation functions---it is not limited to ground state methods and does not require specific measures or susceptibilities. Lastly, let us mention that exploring a connection to alternative notions of entanglement~\cite{Balasubramanian2012,Costa2022,Doi2023,Doi2023a, Grieninger2024,Castro2026,Maldacena1999,Swingle2012} and the recently introduced information lattice~\cite{Kvorning2022,Artiaco2024,Artiaco2025, Bauer2025} that has been applied to the study of metal-insulator transitions~\cite{Skoglund2026} offers an exciting perspective for future work.

\begin{acknowledgments}
\section{Acknowledgements}
We thank Samuel Badr, Gabriele Bellomia, Jan von Delft, Herbert Eßl, Markus Frankenbach, Eric Jacob, Matthias Reitner, Marc Ritter and Nepomuk Ritz for insightful discussions. This work was funded in part  by the Austrian Science Fund (FWF) projects through Grant DOI 10.55776/P36332,  10.55776/F86,  10.55776/I5868, 10.55776/V1018, and 10.55776/PIN4372024. For open access purposes, the authors have applied a CC BY public copyright license to any author-accepted manuscript version arising from this submission. Calculations have been partly performed using Austrian Scientific Computing (ASC).
H.S. was supported by JSPS KAKENHI Grants No. 21H01041, No. 21H01003, and No. 23H03817, JSPS Bilateral Program No. JPJSBP120252002 as well as  JST FOREST Grant No. JPMJFR2232, Japan.
We acknowledge the use of large language models (LLMs) for assistance in the preparation of this manuscript, particularly for text editing and language refinement.
\end{acknowledgments}

\bibliography{Bibliography_QTTdiagnostics}

\clearpage
\appendix

\onecolumngrid
\noindent
\centerline{\textbf{\large End Matter}}
\twocolumngrid

\section{Appendix A: QTTD and multi-point correlators}
\begin{figure}
    \setlength{\abovecaptionskip}{0pt}
    \setlength{\belowcaptionskip}{-14.5pt}
    \centering 
    \includegraphics[width=\linewidth]{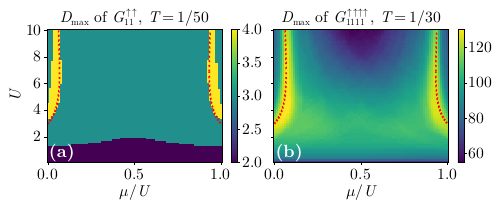}
    \caption{QTT bond dimension \Dmax{} of different imaginary time correlation functions for the Hubbard dimer as a function of electronic repulsion $U$ and chemical potential $\mu/U$: (a)  \Dmax{} for the one-particle Green's function $G_{11}^{\uparrow \uparrow}$ at temperature $T=\beta^{-1}=1/50$, and (b) for the two-particle Green's function $G_{1111}^{\uparrow \uparrow \uparrow \uparrow}$ at $T=1/30$. Red dashed lines mark crossings of the ground state of the model. These crossings can be diagnosed as a sharp maximum in \Dmax{}.
    }
    \label{fig:Hubbard-dimer-appendix}
\end{figure}

Let us consider an arbitrary $n$-point correlator function $\mathcal{G}(\boldsymbol{\tau}) = - \langle \mathcal{T} \hat O_1(\tau_1) \hat O_2(\tau_2) ... \hat O_n(0) \rangle$ with $\boldsymbol{\tau} = (\tau_1,\tau_2,...,\tau_{n-1}), \tau_i\in [0,\beta)$ and fermionic or bosonic operators $\hat O_i$. For $\tau_1>\tau_2> \ldots \tau_{n-1}$, we express the correlator in terms of its Lehmann representation:
\begin{align}
    \mathcal{G}(\boldsymbol{\tau}) &= -\frac{1}{Z} \sum_{m_1\ldots m_{n}} e^{-\tau_1(\epsilon_{m_2}-\epsilon_{m_1})} e^{-\tau_2(\epsilon_{m_3}-\epsilon_{m_2})} \cdots e^{-\tau_{n-1}(\epsilon_{m_n}-\epsilon_{m_{n-1}})} \nonumber \\
    &\times e^{-\beta \epsilon_{m_1}}  \langle m_1 | \hat O_1 | m_2 \rangle \langle m_2 | \hat O_2 | m_3 \rangle \cdots \langle m_n|\hat O_n|m_1\rangle
\end{align}
with the eigenstates $|m_i \rangle$ of the Hamiltonian and we set the ground state energy $\epsilon_{\mathrm{GS}}=0$ in the following. Then, in the case of temperature approaching zero ($\beta \rightarrow \infty$) only summands including the ground state $|\mathrm{GS} \rangle$ ($\exists i, i=1,...,n: |m_i \rangle=|\mathrm{GS}\rangle$) will give non-zero contributions, because all other contributions are exponentially suppressed. If the temperature is finite, but very low, also the other summands will contribute, however they will still be heavily suppressed. Therefore, the correlator will still be primarily dominated by terms containing the ground state. E.g. in the case of a three-point correlator, the following terms are dominating
\begin{subequations}
    \begin{align}
        &|m_1\rangle = |\mathrm{GS} \rangle : \\
        &\quad  e^{-\tau_1 \epsilon_{m_2}} e^{-\tau_2(\epsilon_{m_3}-\epsilon_{m_2})} \langle \mathrm{GS} | \hat O_1 | m_2 \rangle \langle m_2 | \hat O_2 | m_3 \rangle \langle m_3 | \hat O_3 | \mathrm{GS} \rangle \nonumber, \\
        &|m_2\rangle = |\mathrm{GS} \rangle : \\
        &\quad e^{-(\beta-\tau_1) \epsilon_{m_1}} e^{-\tau_2 \epsilon_{m_3}} \langle m_1 | \hat O_1 | \mathrm{GS} \rangle \langle \mathrm{GS} | \hat O_2 | m_3 \rangle \langle m_3 | \hat O_3 | m_1 \rangle \nonumber, \\
        &|m_3\rangle = |\mathrm{GS} \rangle : \\
        &\quad e^{-(\beta - \tau_1) \epsilon_{m_1}} e^{-(\tau_1-\tau_2)\epsilon_{m_2}} \langle m_1 | \hat O_1 | m_2 \rangle \langle m_2 | \hat O_2 | \mathrm{GS} \rangle \langle \mathrm{GS} | \hat O_3 | m_1 \rangle \nonumber,
    \end{align}
\end{subequations}
where we already set $\epsilon_{\mathrm{GS}}=0$. Similar to the analysis of Eq.~\eqref{eq:two-point-correlator-Lehmann-zero-T-crossing}, the first summand dominates in the case of small $\tau_1, \tau_2$, the second one in the case of large $\tau_1$ (close to $\beta$) and small $\tau_2$ and the third for large $\tau_1,\tau_2$. In contrast to the case of the two-point correlator, it can be seen that apart from the dependence on two instead of one imaginary times, in general, more exponentials will contribute in the three-point correlator. Since the QTT representation of an exponential has a bond dimension of one and the bond dimension of the sum of two QTTs is bound by the sum of the bond dimensions of the individual QTTs, we expect  three-point correlators to have larger QTT bond dimensions than two-point correlators due to additional imaginary time dependence and more dominating contributions in the sum. Therefore, higher-point correlators are also expected to show more pronounced peaks at phase transitions and crossovers (cf. Fig.~\ref{fig:Hubbard-dimer} and Fig.~\ref{fig:Hubbard-dimer-appendix}). The reasoning that phase transitions and crossovers are connected to maxima in the bond dimensions of correlators follows the same arguments as in the main text. Let us emphasize that this derivation is valid for arbitrary (non-zero) correlators. Since Fourier transformation can be represented by a low-rank matrix product operator~\cite{Shinaoka2023,Chen2023} applied to the correlator QTT, imaginary frequency counterparts to the imaginary times correlator are expected to display the same maxima in the bond dimension, which is supported by the results in the SM~\cite{SupplementalMaterial}. Hence, the notion of time and frequency scale entanglement will be used interchangeably. Therefore, arbitrary correlators that are not entirely orthogonal to the response of the system and are available from computations can be used in QTTD to diagnose ground state crossings in systems. Following the discussion in the main text, we hypothesize that signatures of enhanced imaginary time-scale entanglement also imprint themselves on generic observables in phase transitions and crossovers resulting from a system-inherent enhancement in time-scale entanglement. 

\section{Appendix B: Diagnostics of ground state crossings }
In the main text, this technique was coined ``quantics tensor train diagnostics''. Here, let us elaborate on the ``diagnostics'' character of the approach for gapped systems (insulating or finite systems) at $T=0$. For this reason, we will revisit Eq.~\eqref{eq:two-point-correlator-Lehmann-zero-T-crossing}. Then, as discussed, the number of exponentials in the sum depends on the matrix elements $\langle \mathrm{GS}_i | \hat A | m \rangle \langle m | \hat B  | \mathrm{GS}_i \rangle$ and $\langle m | \hat A | \mathrm{GS}_i \rangle \langle \mathrm{GS}_i | \hat B  | m \rangle$ being nonzero. However, within one ``phase'' (ground state regime) in parameter space, the number of nonzero terms is fixed because gapped systems are considered without continuous ground state changes. This results in the same number of exponentials contributing to the correlation function, which we denote by $\mathcal{N}_1$. Although, individual contributions might become increasingly suppressed within one ground state regime, they will not entirely vanish leading to a fixed bond dimension matching the number of exponentials in the exact analytical construction of the corresponding QTT. Considering a neighboring but differing phase, the same argument is valid, resulting in the contribution of $\mathcal{N}_2$ exponentials, leading to a bond dimension of $\mathcal{N}_2$. Then a maximum in the bond dimension bound by $\mathcal{N}_1+\mathcal{N}_2$ from above is observed precisely at the crossing of the two ground states, where the different exponentials from both ground states contribute. If the two ground states do not have any matching states $|m\rangle$ that lead to nonzero contributions, the bond dimension in the exact analytical construction at the boundary will be precisely $\mathcal{N}_1+\mathcal{N}_2$. Let us mention that in both phases it might also be possible to see small dips in \Dmax, if two excited states $|m\rangle$ that both lead to nonzero contributions cross, because the contributing exponentials of both states will match each other. Hence, at $T=0$ a maximum in the bond dimension of the exact analytical QTT construction of an arbitrary correlator corresponds to a crossing of two ground state regimes. However, at finite temperatures, all states will contribute to the correlator, which would lead to a constant bond dimension in the entire parameter space in the exact analytical construction. Since at low temperatures most of the contributions will be heavily suppressed, they will be truncated in a numerical QTT construction of the correlator, restoring the diagnostics power of the presented method. However, truncation can come at the price of varying bond dimensions within one ground state regime, which can lead to broader and smaller maxima, also within one ground state regime. Since the system-inherent amount of length and time-scale entanglement is the driving force of the maximum of the bond dimension at a ground state crossing, by studying various correlators and varying the QTT cutoffs and also the temperature, the diagnostics power can be fully restored, where clear stable sharp peaks can be associated with ground state crossings, while broader temperature-dependent peaks are corresponding to thermal crossings.

\section{Appendix C: QTTD for realistic applications }
In contrast to many conventional methods of identifying phase transitions and crossovers, the choice of the QTT cutoff allows for flexibility in dealing with low-accuracy data and, additionally, probing features of scale entanglement, like the imprinting of scale invariance onto generic observables. This can be understood in the following way. Let us consider the case of noise in the evaluation of the correlators. Then the bond dimension of a QTT compression with a cutoff or tolerance below the noise level will suffer from artifacts due to the inclusion of the noise in the compressed QTT. Not only will these artifacts be highly dependent on the chosen cutoff, but they will also manifest themselves in the bond dimension structure. If the QTT cutoff is increased above the noise level, the artifacts will vanish and only features in the QTT bond dimensions associated with phase transitions and crossovers will remain. These features are then expected to be stable for a broad range of different cutoffs above the error level clearly indicating the underlying enhancement in the length- and time-scale entanglement of the system. Varying the QTT cutoff allows to access these different features associated with the enhancement of scale entanglement. Especially the investigation of the manifestation of scale invariance becomes feasible since the distinct feature of uniform scale entanglement is independent of the QTT cutoff for true scale invariance. Therefore, it is possible to not only identify low-accuracy input data, but to overcome its limitations by truncating the noise in the QTT compression. Further evidence and support of this property is provided in the SM \cite{SupplementalMaterial}.

\section{Appendix D: Spin and charge susceptibility}
The local spin susceptibility $\chi_S$ is defined in the following way \cite{Wang2015b}
\begin{align}
    \chi_s=\int_0^{\beta}d \tau \langle \hat S_z(\tau) \hat S_z(0) \rangle,
\end{align}
where $\hat S_z=(\hat n_{\uparrow} - \hat n_{\downarrow})/2$ describes the local magnetization on a single site. Similarly, the local charge susceptibility is defined as follows
\begin{align}
    \chi_c=\int_0^{\beta}d \tau \langle \hat n(\tau) \hat n(0) \rangle,
\end{align}
with $\hat n=(\hat n_{\uparrow} + \hat n_{\downarrow})/2$.

\section{Appendix E: Spectral function of $\mathrm{NdNiO}_2$}
\begin{figure}[H]
    \setlength{\abovecaptionskip}{0pt}
    \setlength{\belowcaptionskip}{-13.5pt}
    \centering
    \includegraphics[width=0.8\linewidth]{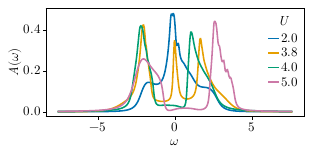}
    \caption{Analytically continued~\cite{ana_cont} spectral function $A(\omega)$ in real frequencies $\omega$ shown for various $U$.}
    \label{fig:spectral_NNO}
\end{figure}

Fig.~\ref{fig:spectral_NNO} shows the analytically continued spectral function for $\mathrm{NdNiO}_2$. It can be seen that spectral weight at $\omega=0$ drops to zero between $U=3.8$ and $4.0$ indicating the emergence of the insulating state.

\section{Appendix F: Transverse-field Ising model} \label{app:tfim}
Following Ref.~\onlinecite{Mbeng2024}, the exact normal Green's function is obtained through a Jordan-Wigner transformation and is defined in the following way
\begin{align}
    G (k,i\nu) :=G_{\mathrm{normal}}(k,i\nu) &= -\frac{i\nu + 2 \left(h-J \cos(k) \right)}{\nu^2 + \epsilon_k^2}, 
\end{align}
with 
\begin{align}
    \epsilon_k = 2J \sqrt{\left(\frac{h}{J} - \cos(k) \right)^2 + \sin(k)^2}.
\end{align}
The local normal Green's function 
\begin{small}
    \begin{align}
    G(i\nu) = -\frac{  \left(\sqrt{\left(4 h^2+4 J^2+\nu ^2\right)^2-64 h^2 J^2}+4 h^2+4 i h \nu -4 J^2-\nu ^2\right)}{4 h \sqrt{\left(4 h^2+4 J^2+\nu ^2\right)^2-64 h^2 J^2}}
\end{align}
\end{small}
is obtained via analytic integration of $G(k,i\nu)$ on the interval $k\in [-\pi,\pi]$.

\end{document}


\preprint{APS/123-QED}

\title{Supplemental Material: Diagnosing phase transitions through time-scale entanglement
}

\author{Stefan Rohshap\,\orcidlink{0009-0007-2953-8831}}
\email{stefan.rohshap@tuwien.ac.at}
\affiliation{%
    Institute of Solid State Physics, TU Wien, 1040 Vienna, Austria}%

\author{Hirone Ishida\,\orcidlink{0009-0000-7095-7105}}
\affiliation{%
    Department of Physics, Saitama University, Saitama 338-8570, Japan}%

\author{Frederic Bippus\,\orcidlink{0009-0006-4316-6547}}
\affiliation{%
    Institute of Solid State Physics, TU Wien, 1040 Vienna, Austria}%

\author{Leonard M. Verhoff\,\orcidlink{0009-0004-3358-1312}}
\affiliation{%
    Institute of Solid State Physics, TU Wien, 1040 Vienna, Austria}%

\author{Anna Kauch\,\orcidlink{0000-0002-7669-0090}}
\affiliation{%
    Institute of Solid State Physics, TU Wien, 1040 Vienna, Austria}%
    
\author{Karsten Held\,\orcidlink{0000-0001-5984-8549}}
\affiliation{%
    Institute of Solid State Physics, TU Wien, 1040 Vienna, Austria}

\author{Hiroshi Shinaoka\,\orcidlink{0000-0002-7058-8765}}
\affiliation{%
    Department of Physics, Saitama University, Saitama 338-8570, Japan}%

\author{Markus Wallerberger\,\orcidlink{0000-0002-9992-1541}}
\affiliation{%
    Institute of Solid State Physics, TU Wien, 1040 Vienna, Austria}%

\date{\today}

\begin{abstract}
    This supplementary material contains additional information to support the conclusions reached in the main text. Section~\ref{sec:entanglement} introduces the fidelity susceptibility and various entanglement measures relevant for the additional results provided in the supplementary material. Section~\ref{sec:dimer} examines the possibility of identifying excited state crossings as well as ground state crossings, further discusses time-scale entanglement to be a system-inherent property that peaks at phase transitions and crossovers becoming visible in the QTT bond dimension and elaborates on how to deal with noisy input data for the Hubbard dimer case. In section~\ref{sec:4-site}, the thermal crossover identified with QTTD in the case of the four-site Hubbard ring with nearest neighbor hopping only is further examined through various entanglement measures and the double occupancy. The case of the four-site Hubbard ring with next-nearest neighbor hopping is further discussed in section~\ref{sec:4-site-nnn} showing an example of a ground state crossing found by QTTD, but failed to be diagnosed by the fidelity susceptibility. Section~\ref{sec:Mott} provides additional results for the cases of the Mott transition in Hubbard model on the Bethe lattice and 3D $\mathrm{NdNiO}_2$. In section~\ref{sec:siam}, details on the computations for SIAM are presented. The TFIM is further discussed in section~\ref{sec:tfim}. Section~\ref{sec:data} refers to the data availability of the publication and section~\ref{sec:computation} outlines computational details of this work.
\end{abstract}
\maketitle

\section{Entanglement measures} \label{sec:entanglement}
\subsection{Fidelity susceptibility \label{sec:fidelity}}
Following Ref.~\onlinecite{Wang2015}, we define the fidelity susceptibility~\cite{Zanardi2006, You2007, Zanardi2007, Venuti2007, Kashihara2023} at finite temperatures in the following way.

Let $\hat H(\lambda)$ be a Hamiltonian depending on the ``driving parameter'' $\lambda$ with
\begin{align}
    \hat H(\lambda) = \hat H_0 + \lambda \hat H_1.
\end{align}
Then, $\lambda$ is chosen so that it parametrizes the competition between $\hat H_0$ and $\hat H_1$ in the way that the system may undergo one or multiple phase transition(s) by varying $\lambda$. $\hat H_0$ will be chosen as the non-interacting part and $\hat H_1$ as the interacting part, with the Coulomb repulsion $U$ as the driving parameter $\lambda$ of the phase transition. The fidelity susceptibility is then defined as
\begin{align}
    \chi_F  = \int_0^{\frac{\beta}{2}} d \tau \left[ \langle \hat H_1(\tau) \hat H_1 \rangle - \langle \hat H_1 \rangle^2 \right] \tau
\end{align} 
with 

\begin{align}
    \hat H_1(\tau) = e^{\hat H \tau} \hat H_1 e^{-\hat H \tau}.
\end{align}
The fidelity susceptibility describes how susceptible the overlap of neighboring ground states (or density matrices in the case of finite temperatures) in parameter space is to small (driving) parameter changes. Similarly to QTTD, the fidelity susceptibility is expected to show maxima at phase transitions. 

\subsection{Various entanglement measures} \label{sec:entanglement-measures}
The quantum Fisher information (QFI) \cite{hauke_measuring_2016}
\begin{equation}
    F_Q\equiv\int_0^{\infty}\tanh\left(\frac{\omega\beta}{2}\right)\textrm{Im}\chi(\omega)
\end{equation}
serves as an experimentally accessible measure of multipartite entanglement. A value $F_Q\geq m$ detects $m+1$-partite entanglement. 
A peak in the derivative of the QFI hints at a phase transition \cite{Qu_phase_transition_FQ,Yamashika2025}.

The two-site mutual information is a measure of total correlations between two lattice sites based on the von Neumann entropy\cite{zurek_information_1983,barnett_entropy_1989}
\begin{equation}
    I_{ij} \equiv \textrm{tr}[\rho_{ij}\ln\rho_{ij}] - 2\textrm{tr}[\rho_i\ln\rho_i],
\end{equation}
where $\rho_{ij}$ is the two-site reduced density matrix and $\rho_i = \rho_j = \textrm{tr}_i[\rho_{ij}]$ \cite{roosz_two-site_2024, Bippus2025a, Bippus2025b}.
This can be generalized to the Rényi mutual information \cite{grover_entanglement_2013,renyi_measures_1961}
\begin{equation}
    I_R \equiv \ln \left[\textrm{tr}  \rho_{ij}^2  \right] - 2 \ln \left[\textrm{tr} \rho_{i}^2  \right].
\end{equation}

The partial transpose $\rho_{ij}^{T_j}$ of the two-site reduced density matrix provides another entanglement measure \cite{peres_separability_1996,horodecki_separability_1996}. If the density matrix is separable, then $\rho_{ij}^{T_j}=\rho_i\otimes\rho_j^T$ is a physical density matrix that is positive semi-definite. Vice versa, negative eigenvalues $\lambda$ of $\rho_{ij}^{T_j}$ indicate entanglement and the negativity is defined as
\begin{equation}
    N\equiv\sum_{\lambda<0}|\lambda|.
\end{equation}

In a phase transition, the density matrix undergoes a rapid change. We illustrate this with a trivial example, in a zero temperature transition from a paramagnetic to a ferromagnetic state, all antiferromagnetic contributions vanish and the density matrix has a discontinuous jump. At higher temperatures this will be less sharp as excited states may still contribute. Similarly, in a crossover, a smooth change between the density matrices of the two regimes is expected. 
Since the two-site mutual information and the negativity are directly computed from the density matrix, such a change shows as an extrema in the derivatives of these measures as well \cite{DeChiara_2018,ZHENG_negativity_phase_transition,Das_I_phase_transition}.
For example, the Mott transition of the Hubbard model shows a clear signature both in the mutual information and in the negativity \cite{Bellomia_entanglment_MI,bellomia2025localclassicalcorrelationsphysical}.

\section{Hubbard dimer \label{sec:dimer}}
\begin{figure}
    \centering 
    \includegraphics[width=\linewidth]{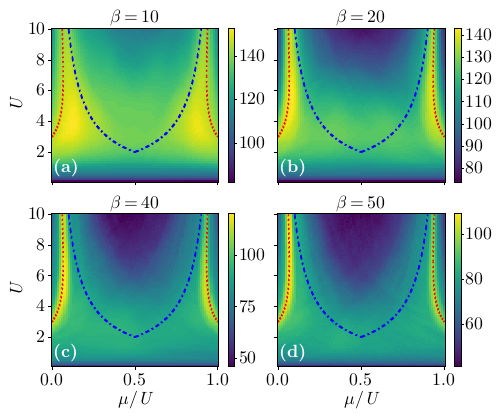}
    \caption{Hubbard dimer: \Dmax{} of the QTT of $G_{1111}^{\uparrow \uparrow \uparrow \uparrow}(\tau_1, \tau_2, \tau_3)$ at $\epsilon=10^{-14}$ for various temperatures. The red dashed lines indicate a GS crossings (singlet - doublet), while the blue dashed line represents a crossing of the first excited state, which also becomes visible in \Dmax. Lowering the temperature leads to localization of the peaks of the QTT bond dimension at the GS crossings.}
    \label{fig:Hubbard-dimer-temperature}
\end{figure}

\begin{figure}
    \centering 
    \includegraphics[width=\linewidth]{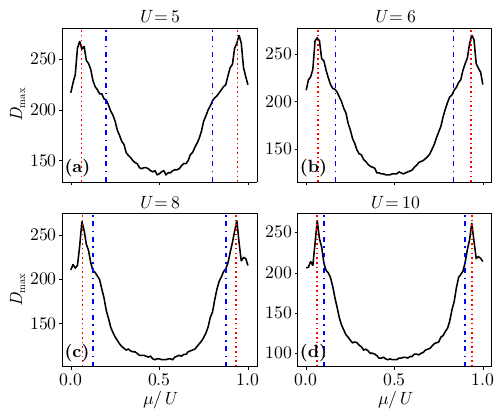}
    \caption{Hubbard dimer: $U$ slices of the \Dmax{} of the QTT of $G_{1111}^{\uparrow \uparrow \uparrow \uparrow}(\tau_1, \tau_2, \tau_3)$ at $\epsilon=10^{-20}$ at $\beta=50$. The red dashed lines indicate a GS crossings (singlet - doublet), while the blue dashed line represents a crossing of the first excited state. The first excited state crossings in the eigenspectrum can be seen as kinks in the QTT bond dimension.}
    \label{fig:Hubbard-dimer-U-slices}
\end{figure}

In this section, we will show additional results of the QTTD approach applied to the case of the Hubbard dimer that might help to further understand the power of the QTTD approach. In Fig.~\ref{fig:Hubbard-dimer-temperature}, the bond dimension of the QTT compression of the four-point correlator $G_{1111}^{\uparrow \uparrow \uparrow \uparrow}(\tau_1, \tau_2, \tau_3) = - \langle \mathcal{T} \hat c_{1,\uparrow} (\tau_1) \hat c_{1,\uparrow}^{\dagger} (\tau_2) \hat c_{1,\uparrow} (\tau_3) \hat c_{1,\uparrow}^{\dagger} \rangle$ is shown for various temperatures. The red dotted line indicates a singlet-doublet ground state transition. Furthermore, the blue dash-dotted line indicates a crossing of the first excited state. At higher temperatures the maximum of the QTT bond dimensions smears out, while at low temperatures (high $\beta$) the peaks lie precisely on top of the ground state crossings. Interestingly, also fingerprints of the crossing of the first excited state can be seen since the bond dimensions also seem to follow the corresponding parabolic shape. 
This finding is further supported in Fig.~\ref{fig:Hubbard-dimer-U-slices}, where different $U$-slices of the bond dimension of $G_{1111}^{\uparrow \uparrow \uparrow \uparrow}(\tau_1, \tau_2, \tau_3)$ at $\beta=50$ are shown. In addition to the sharp peaks in \Dmax{} corresponding to QPTs (red dotted lines), fingerprints of the crossings of the first excited states become also visible in the form of kinks in the bond dimension in the vicinity of the associated points. 

\begin{figure}
    \centering 
    \includegraphics[width=\linewidth]{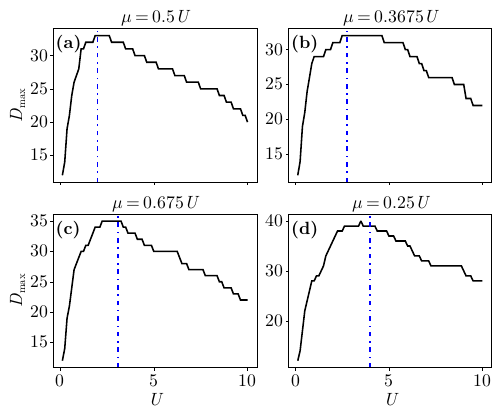}
    \caption{Hubbard dimer: $\mu$ slice \Dmax{} of the QTT of $G_{1111}^{\uparrow \uparrow \uparrow \uparrow}(\tau_1, \tau_2, \tau_3)$ computed with a truncated eigenspectrum at $\epsilon=10^{-8}$ and $\beta=50$. The blue dashed line represents a crossing of the first excited state, which can also be seen as maxima in \Dmax.}
    \label{fig:Hubbard-dimer-mu-slices}
\end{figure}

Additionally, when examining different $\mu$-slices of the QTT bond dimension of $G_{1111}^{\uparrow \uparrow \uparrow \uparrow}(\tau_1, \tau_2, \tau_3)$ (with truncated eigenspectrum) in Fig.~\ref{fig:Hubbard-dimer-mu-slices}, crossings of the first excited states can even be seen as peaks in \Dmax{} of the QTT. However, let us mention that the precise position of these maxima and the kinks in Fig.~\ref{fig:Hubbard-dimer-U-slices} are more susceptible to temperature and the choice of the QTT cutoff, while the peaks associated with ground state crossings are very stable. Again, this shows the potential of the newly developed QTTD approach and we hope that these results will spark further research and ideas on how to identify excited state crossings with QTTD. 

\begin{figure}
    \centering 
    \includegraphics[width=\linewidth]{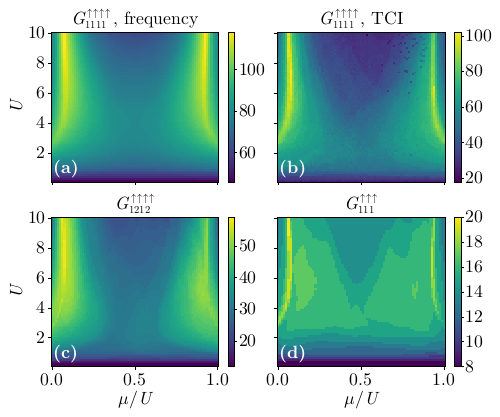}
    \caption{Hubbard dimer: \Dmax{} of the QTT of various correlators computed with a truncated eigenspectrum at $\epsilon=10^{-8}$ and $\beta=50$. In (a), the \Dmax{} of the Fourier transform of $G_{1111}^{\uparrow \uparrow \uparrow \uparrow}(\tau_1, \tau_2, \tau_3)$ to Matsubara frequencies is presented, while in (c) and (d) the imaginary times four- and three-point correlators $G_{1212}^{\uparrow \uparrow \uparrow \uparrow}(\tau_1, \tau_2, \tau_3)$ and $G_{111}^{\uparrow \uparrow \uparrow}(\tau_1, \tau_2)$ are shown. In (b), the bond dimension of the QTT of $G_{1111}^{\uparrow \uparrow \uparrow \uparrow}(\tau_1, \tau_2, \tau_3)$ compressed with TCI to a tolerance (maximum norm) of $\epsilon=10^{-5}$ is displayed.}
    \label{fig:Hubbard-dimer-various}
\end{figure}

In the next step, we provide further evidence for system-inherent time-scale entanglement as a core feature of QTTD. In the derivation of the QTTD method in the main text and in App. A it was emphasized that the derivation is valid for arbitrary correlators, meaning that the correlators show the same peaks associated with phase transitions and thermal crossovers. In Fig.~\ref{fig:Hubbard-dimer-various}, the QTT bond dimension of different correlators (with eigenspectrum truncation) is presented in the $\mu/U-U$-plane at $\beta=50$. In Fig.~\ref{fig:Hubbard-dimer-various}(a), \Dmax{} of the Fourier transform of $G_{1111}^{\uparrow \uparrow \uparrow \uparrow}(\tau_1, \tau_2, \tau_3)$ to Matsubara frequencies is shown supporting the conjecture of the main text that the same features can be identified using Matsubara frequency correlators. This is due to the fact that the Fourier transform to Matsubara frequencies can be represented by a low rank matrix product operator applied to the QTT of the imaginary times correlator, which, hence, maps regions with large bond dimensions again to the same parameters.
In Figs.~\ref{fig:Hubbard-dimer-various}(c) and (d), we present the four- and three point correlators $G_{1212}^{\uparrow \uparrow \uparrow \uparrow}(\tau_1, \tau_2, \tau_3) = - \langle \mathcal{T} \hat c_{1,\uparrow} (\tau_1) \hat c_{2,\uparrow}^{\dagger} (\tau_2) \hat c_{1,\uparrow} (\tau_3) \hat c_{2,\uparrow}^{\dagger} \rangle$ and $G_{111}^{\uparrow \uparrow  \uparrow}(\tau_1, \tau_2) = - \langle \mathcal{T} \hat c_{1,\uparrow} (\tau_1) \hat n_{1,\uparrow} (\tau_2)  \hat c_{1,\uparrow}^{\dagger} \rangle$. Although the eigenspectrum was truncated in the evaluation of the correlators and the SVD cutoff is set to $\epsilon=10^{-8}$, it can still clearly be seen that all correlators precisely show the same peaks in their bond dimensions, where time-scale entanglement in the system is increased due to the occurrance of a QPT. Furthermore, in~\ref{fig:Hubbard-dimer-various}(b), these features are shown to also be stable using tensor cross interpolation (TCI) as compression method in comparison to SVD applied in the other cases. These results support the notion of system-inherent time-scale entanglement as a key finding emphasizing the flexibility of the approach due to the possibility of using arbitrary correlators in QTTD. This contrasts with most conventional methods, where specific order parameters or susceptibilities have to be studied in detail that might not be easily accessible. With QTTD on the other hand, phase transitions and crossovers can be diagnosed using correlators that are easily accessible. 

\begin{figure}
    \centering 
    \includegraphics[width=\linewidth]{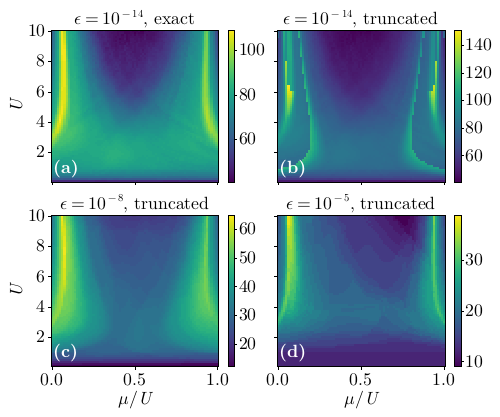}
    \caption{Hubbard dimer: \Dmax{} of the QTT of $G_{1111}^{\uparrow \uparrow \uparrow \uparrow}(\tau_1, \tau_2, \tau_3)$ at $\beta=50$. In (a), the result is shown for the computation of the correlator with the entire eigenspectrum at $\epsilon=10^{-14}$. Artifacts in the bond dimension emerge at the same QTT cutoff due to ``noise'' (truncation of eigenspectrum) in (b). In (c) and (d) these artifacts can be removed by increasing the QTT cutoff above the noise level to $\epsilon=10^{-8}$ and $\epsilon=10^{-5}$ showing stable peaks of the QTT bond dimension at the ground state crossing.}
    \label{fig:Hubbard-dimer-cutoffs}
\end{figure}

Besides relying on the computation of specific properties, other methods of probing phase transitions and crossovers can also be severely limited by the accuracy of the underlying method producing a single ``best effort'' dataset from which conclusions are drawn. In contrast, using QTTD we can vary the cutoff or tolerance in the QTT compression a posteriori and by this thoroughly examine the available data and, hence, deal with low-accuracy data as briefly discussed in App. B. Fig.~\ref{fig:Hubbard-dimer-cutoffs}, illustrates this aspect of the QTTD method, where the QTT bond dimension of $G_{1111}^{\uparrow \uparrow \uparrow \uparrow}(\tau_1, \tau_2, \tau_3)$ is presented in the $\mu/U-U$-plane at $\beta=50$. In Fig.~\ref{fig:Hubbard-dimer-cutoffs}(a), we show the same result as in the discussion of the Hubbard dimer in the main text. Here, the eigenspectrum is not truncated in the calculation of the correlators and the SVD cutoff is set to $\epsilon=10^{-14}$. Since in realistic computations the entire eigenspectrum will not be accessible in an exact manner and is not necessary for the QTTD method relying on arbitrary correlators instead, in Figs.~\ref{fig:Hubbard-dimer-cutoffs} (b)-(d) the eigenspectrum is truncated. In the case of a QTT cutoff of $\epsilon=10^{-14}$ (see (b)), it can be seen that some artifacts occur due to the truncation of the eigenspectrum and, hence, the lower accuracy in the data. However, the position of these artifacts heavily depends on the cutoff and they can be removed by increasing the QTT cutoff as can be seen in (c) ($\epsilon=10^{-8}$) and (d) ($\epsilon=10^{-5}$). In both cases the low-accuracy artifacts have been removed resulting in stable maxima in the QTT bond dimensions across a wide range of cutoffs. This can be explained in the following way. The trunctation of the eigenspectrum can be interpreted as introducing noise in the data. Therefore, choosing a too low QTT cutoff leads to bond dimension artifacts emerging in Fig.~\ref{fig:Hubbard-dimer-cutoffs} (b), since the ``noise'' is not truncated but included in the bond dimension. By increasing the QTT cutoff in (c) and (d) above the ``noise'' level, the artifacts are removed since the noise is truncated in the QTT compression. Therefore, in this way also noisy data can be dealt with by increasing the QTT cutoff until stable peaks emerge indicating phase transitions and crossovers.  

\begin{figure}
    \centering 
    \includegraphics[width=\linewidth]{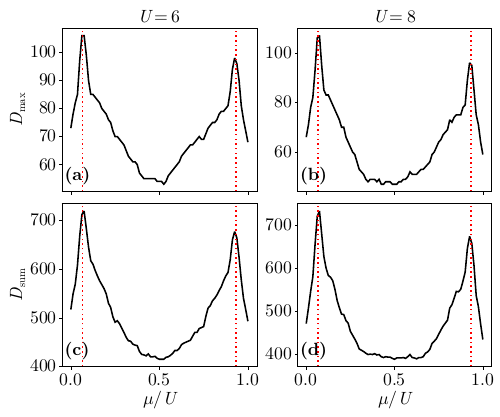}
    \caption{Hubbard dimer: \Dmax{} ((a),(b)) vs. \Dsum{} ((c),(d)) of the QTT of $G_{1111}^{\uparrow \uparrow \uparrow \uparrow}(\tau_1, \tau_2, \tau_3)$ at $\beta=50, \epsilon=10^{-14}$ for $U=6$ ((a),(c)) and $U=8$ ((b),(d)). Red dashed lines indicate point of ground state crossing.}
    \label{fig:Hubbard-dimer-maxdim-vs-sumdim}
\end{figure}

Finally let us show that in the case of the Hubbard dimer \Dsum{} and \Dmax{} show the same peaks as can be seen in Fig.~\ref{fig:Hubbard-dimer-maxdim-vs-sumdim}, where the two measures of scale entanglement are shown for $U=6$ and $8$ as a function of the chemical potential. Therefore, for the ground state crossing of the Hubbard dimer the total entanglement between all time scales measured by \Dsum{} and the maximum amount of entanglement between two time scales (\Dmax{}) show the same qualitative behavior.

\section{4-site Hubbard ring \label{sec:4-site}}
We consider the periodic 1d 4-site $(N=4)$ Hubbard ring with nearest neighbor hopping only, i.e., $N=4, t' = 0$, $t_{\pm(N-1)}=t$, in the following Hubbard Hamiltonian
\begin{equation}
    \hat H = -\sum_\sigma \sum_{i,j=1}^N t_{i-j} \hat c_{i,\sigma}^{\dagger} \hat c_{j,\sigma} 
    + \sum_{i=1}^N \left( U  \hat n_{i, \uparrow} \hat n_{i,\downarrow} - \mu (\hat n_{i,\uparrow} + \hat n_{i,\downarrow}) \right).
     \label{eq:general-Hubbard-Hamiltonian}
\end{equation}

\begin{figure}
    \setlength{\abovecaptionskip}{0pt}
    \setlength{\belowcaptionskip}{-5.5pt}
    \centering 
    \includegraphics[width=\linewidth]{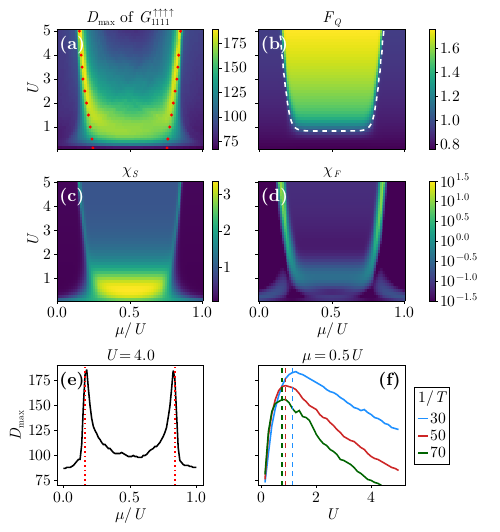}
    \caption{Four-site Hubbard ring with nearest-neighbor hopping only: (a) \Dmax{} of the QTT for $G_{1111}^{\uparrow \uparrow \uparrow \uparrow}(\tau_1, \tau_2, \tau_3)$ at $1/T=\beta=50$, $\epsilon=10^{-14}$ and $R=6$ in comparison to (b) the quantum Fisher information $F_Q$, where the white dashed line indicates $F_Q=1$
    , (c)  spin susceptibility $\chi_S$ and (d) fidelity susceptibility. (e) and (f):  $U=4$ and $\mu=U/2$ slice of (a), where red dashed lines {in (e)}  indicate QPTs {and dashed} lines in (f) indicate $F_Q=1$. In (f), two additional $\beta$'s are shown.}
    \label{fig:four-site-QTTD}
\end{figure}

In Fig.~\ref{fig:four-site-QTTD}(a), 
\Dmax{} of the QTT representation of the two-particle Green's function $G_{1111}^{\uparrow \uparrow \uparrow \uparrow}(\tau_1, \tau_2, \tau_3)$ is shown in the $\mu/U-U$-plane. Crossings of the ground states are indicated by red dots. For not too low values of $U$ 
the dots lie precisely on top of the peaks of \Dmax{}. 
In Fig.~\ref{fig:four-site-QTTD}(e), the slice for $U=4$ is shown.
For lower values of $U$ 
the \Dmax{} peaks deviate from 
the crossings in the eigenspectrum of the Hamiltonian (red dashed lines) and become broader. 
This is because for small $U$ and $\beta=50$ excited states become thermally accessible.
Consequently, when decreasing $T$, 
the region with large \Dmax{} shifts to smaller $U$'s, see Fig.~\ref{fig:four-site-QTTD}(f), and for $T\rightarrow 0$ again alienates with the eigenvalue crossings.

The spin susceptibility $\chi_S$ in Fig.~\ref{fig:four-site-QTTD}(c) and, e.g., the fidelity susceptibility $\chi_F$ (in Fig.~\ref{fig:four-site-QTTD}(d) 
(definition in
Sec.~\ref{sec:fidelity}) show a qualitatively similar behavior as \Dmax{}: at large values of $U$ the ground state crossings are marked by a maximum, while for small values of $U$ both show larger values, without an eigenvalue crossing.
Further, when investigating entanglement in Fig.~\ref{fig:four-site-QTTD}(b), also a similar structure is seen, where the white dashed line indicates the quantum Fisher information $F_Q=1$ 
(definition in Sec.~\ref{sec:entanglement-measures}). Hence, the peaks in \Dmax{} for low $U$ in Fig.~\ref{fig:four-site-QTTD}(a) seem to be associated with some thermal crossover between different regimes. Further evidence of this thermal crossover seen by QTTD 
is seen in Fig.~\ref{fig:four-site-QTTD}(f), where the QTT bond dimension of $G_{1111}^{\uparrow \uparrow \uparrow \uparrow}$ is shown for various temperatures. The dashed lines indicate  $F_Q=1$ and are located at the maxima of the different curves. Interestingly, also the derivative 
$\frac{d F_Q}{dU}$ (and other entanglement measures  as well as that of the double occupancy
show a maximum in close vicinity of this point. Additionally, also the maximum and inflection point of the spin susceptibility lie in proximity to this point and show a similar temperature behavior. We will further investigate these properties in the following.

\begin{figure}
    \centering 
    \includegraphics[width=\linewidth]{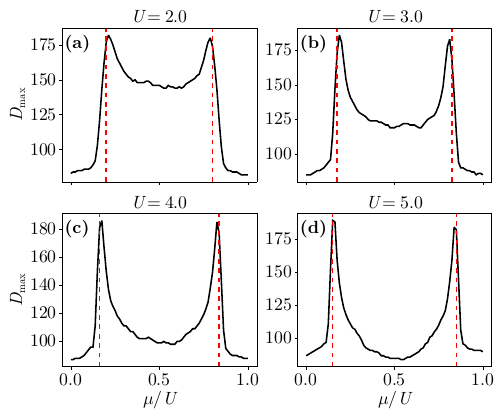}
    \caption{Four-site Hubbard ring with nearest-neighbor hopping only: $U$ slices of the \Dmax{} of the QTT of $G_{1111}^{\uparrow \uparrow \uparrow \uparrow}(\tau_1, \tau_2, \tau_3)$ at $\epsilon=10^{-14}$ at $\beta=50$. The red dashed lines indicate a GS crossings.}
    \label{fig:four-site-U-slices-QTTD}
\end{figure}

In Fig.~\ref{fig:four-site-U-slices-QTTD}, different U slices of the QTT bond dimension of the four-point correlator $G_{1111}^{\uparrow \uparrow \uparrow \uparrow}(\tau_1, \tau_2, \tau_3)$ (compare with Fig. 3(a) in the main text) are shown. It can be seen that the sharp peaks in the maximum bond dimension are located at the red dashed lines that indicate a QPT. 

\begin{figure}
    \centering 
    \includegraphics[width=\linewidth]{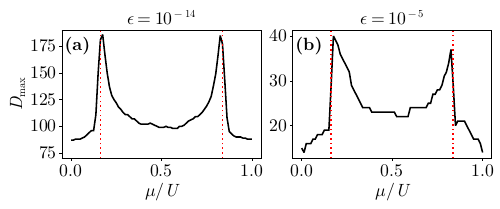}
    \caption{Four-site Hubbard ring with nearest-neighbor hopping only: \Dmax{} of the QTT of $G_{1111}^{\uparrow \uparrow \uparrow \uparrow}(\tau_1, \tau_2, \tau_3)$ at $\epsilon=10^{-14}$ at $\beta=50, U=4$ for QTT cutoff $\epsilon=10^{-14}$ (a) and $\epsilon=10^{-5}$ (b). The red dashed lines indicate a GS crossings.}
    \label{fig:four-site-cutoffs}
\end{figure}

In Fig.~\ref{fig:four-site-cutoffs} the QTT maximum bond dimension for the $U=4$ slice is shown for two different QTT cutoffs $\epsilon=10^{-14}$ and $10^{-5}$. It can be seen that the \Dmax{} curves are not smooth due to the numerical QTT compression. Still, since the curves are obtained via compression with SVD, they are expected to be smoother than TCI-compressed ones.

\begin{figure}
    \centering 
    \includegraphics[width=\linewidth]{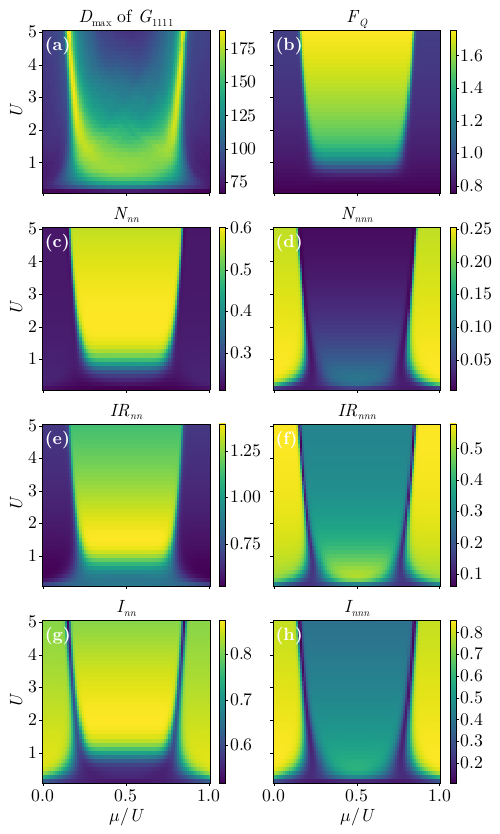}
    \caption{Various entanglement measures in comparison to the \Dmax{} of the QTT of $G_{1111}^{\uparrow \uparrow \uparrow \uparrow}(\tau_1, \tau_2, \tau_3)$ at $\epsilon=10^{-14}$ at $\beta=50$ for the four-site Hubbard ring with nearest neighbor hopping only. (b) presents the quantum Fisher information $F_Q$, while (c), (e) and (g) show the nearest-neighbor negativity ($N_{nn}$), Renyi mutual information ($IR_{nn}$) and mutual information ($I_{nn}$). (d), (f) and (h) present the corresponding next-nearest neighbor measures.}
    \label{fig:four-site-entanglement-QTTD}
\end{figure}

In Fig.~\ref{fig:four-site-entanglement-QTTD}, the QTT bond dimension of $G_{1111}^{\uparrow \uparrow \uparrow \uparrow}(\tau_1, \tau_2, \tau_3)$ (a) is shown in comparison to various entanglement measures at $\beta=50$. Already here it can be seen that QTTD diagnoses different ``regimes'' of entanglement meaning rapid changes in different entanglement measures in parameter space due to phase transitions and crossovers. For example, the quantum Fisher information $F_Q$ in (b) becomes one indicating the start of entanglement precisely at the maximum of the bond dimensions in (a). In Figs. 3 (b),(f) in the main text, it was shown that the maximum in \Dmax{} corresponds to $F_Q=1$ also for different temperatures. The nearest-neighbor negativity $N_{nn}$ in (c) and the nearest-neighbor Renyi mutual information $IR_{nn}$ in (e) have a very similar behavior. Apart from diagnosing the rapid changes in entanglement, it is interesting that in comparing the QTT bond dimension to the nearest-neighbor mutual information $I_{nn}$ in (g) it looks as if $I_{nn}$ was the photo negative of the \Dmax{} in (a).

\begin{figure}
    \centering 
    \includegraphics[width=\linewidth]{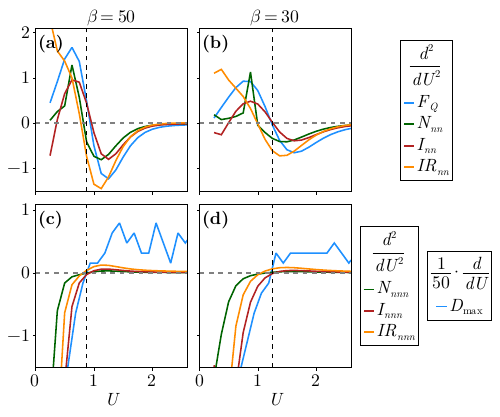}
    \caption{Four-site Hubbard ring with nearest-neighbor hopping only: Second derivatives of the various entanglement measures with respect to $U$ for $\beta=50$ (a), (c) and $\beta=30$ (b), (d). (a) and (b) show the nearest-neighbor negativity $N$, Renyi mutual information $IR$ and mutual information $I$ as well as the quantum Fisher information $F_Q$. (c) and (d) show the corresponding next-nearest neighbor measures as well as the first derivative of the QTT bond dimension of $G_{1111}^{\uparrow \uparrow \uparrow \uparrow}(\tau_1, \tau_2, \tau_3)$, where the black dashed line indicates a zero (maximum of \Dmax).}
    \label{fig:four-site-entanglement_derivatives-QTTD}
\end{figure}

Let us now focus on the thermal phase crossover discussed in Fig. 3(f) in the main text. Here, a broad maximum in the QTT bond dimension of  $G_{1111}^{\uparrow \uparrow \uparrow \uparrow}(\tau_1, \tau_2, \tau_3)$ for various temperatures was located at the point, where the quantum Fisher information takes the value one (start of entanglement). In Fig.~\ref{fig:four-site-entanglement_derivatives-QTTD}, we support the finding of thermal phase crossover with QTTD by examining the second derivatives of the different entanglement measures for $\beta=30,50$. The rescaled first derivative of the QTT bond dimension of the four-point correlator is shown as the blue line in (c) and (d) and the dashed black line indicates, where the maximum of the QTT \Dmax{} is located. It can be seen that the second derivatives of the different nearest-neighbor (nn), next-nearest neighbor (nnn) entanglement measures as well as the quantum Fisher information cross zero in close vicinity of the maximum of the QTT \Dmax{} indicating an inflection point in all of these functions. The inflection point in all of the entanglement measures provides further evidence for a thermal phase crossover at this point that was discovered with the help of QTTD. 

\begin{figure}
    \centering 
    \includegraphics[width=\linewidth]{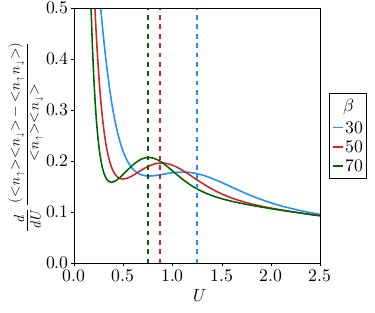}
    \caption{Four-site Hubbard ring with nearest-neighbor hopping only: First derivative of the double occupancy $\left( \langle \hat n_{\uparrow} \rangle \langle \hat n_{\downarrow} \rangle - \langle \hat n_{\uparrow} \hat n_{\downarrow} \rangle \right)/\left( \langle \hat n_{\uparrow} \rangle\langle \hat n_{\downarrow} \rangle \right)$ with respect to $U$ for various temperatures, where the dashed lines indicate the corresponding maxima of the QTT bond dimension of $G_{1111}^{\uparrow \uparrow \uparrow \uparrow}(\tau_1, \tau_2, \tau_3)$.}
    \label{fig:four-site-double-occupancy-QTTD}
\end{figure}

In Fig.~\ref{fig:four-site-double-occupancy-QTTD}, further evidence and insights into the nature of this thermal phase crossover are provided. Here, the first derivative of the double occupancy $\left( \langle \hat n_{\uparrow} \rangle \langle \hat n_{\downarrow} \rangle - \langle \hat n_{\uparrow} \hat n_{\downarrow} \rangle \right)/\left( \langle \hat n_{\uparrow} \rangle\langle \hat n_{\downarrow} \rangle \right)$ is shown. The double occupancy itself is zero at $U=0$ and approaches one for larger $U$. In infinite systems a value of zero would correspond to a metallic state, while a value of one would be an insulating state. Additionally, a jump or rapid change in the double occupancy can indicate a Mott metal-insulator transition. In Fig.~\ref{fig:four-site-double-occupancy-QTTD}, it can be seen that the local maxima of the first derivative of the double occupancy indicating inflection points and, thus, rapid changes of the double occupancy are located close to the maxima of the QTT bond dimension of the four-point correlator (indicated by the dashed lines) for various bond dimensions. Since the four-site Hubbard ring is a small finite-sized system, one has to be careful in interpreting the results as a Mott metal-insulator transition. However, in contrast to e.g. the Hubbard dimer, the four-site Hubbard ring has a degenerate ground state at $U=0$ that splits into a singlet ground state and a triplet excited state for finite $U$. With finite temperature this ``metal-insulator-like'' crossover is pushed to higher values of $U$, which again matches the observed results. 
Since this crossover was unknown to the authors, the maximum of the QTT bond dimension associated with this crossover was surprising, which is why the broad set of discussed indicators was then studied clearly supporting the finding of a thermal phase crossover with QTTD.

\section{4-site Hubbard ring with $t'$ \label{sec:4-site-nnn}}
\begin{figure}
    \setlength{\abovecaptionskip}{0pt}
    \setlength{\belowcaptionskip}{-13.5pt}
    \centering 
    \includegraphics[width=\linewidth]{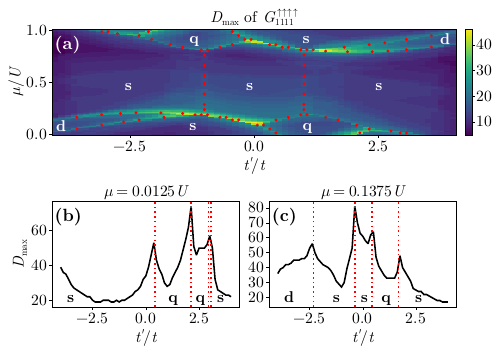}
    \caption{Four-site Hubbard ring with next nearest-neighbor hopping:  Red dots (dotted lines) indicate crossings of s(inglet), d(oublet) and q(uartet) ground states. (a) \Dmax{} of QTT of $G_{1111}^{\uparrow \uparrow \uparrow \uparrow}(\tau_1,\tau_2,\tau_3)$ at $\beta=50$ with $\epsilon=10^{-5}$ and $R=6$. (b)-(c) \Dmax{} for various $\mu/U$-slices at $\epsilon=10^{-8}$.}
    \label{fig:four-site-next-nearest-QTTD}
\end{figure}

We consider \Dmax{} of 
$G_{1111}^{\uparrow \uparrow \uparrow \uparrow}$ of the periodic 4-site Hubbard ring with next-nearest neighbor hopping $t'$ at $ U=10$, $\beta=50$
in the $t'-\mu/U$-plane \cite{Nishimoto2008}. Contrary to the previous examples, we apply larger QTT cutoffs $\epsilon$. In Fig. \ref{fig:four-site-next-nearest-QTTD}(a), \Dmax{} of $G_{1111}^{\uparrow \uparrow \uparrow \uparrow}(\tau_1,\tau_2,\tau_3)$ is shown for $\epsilon=10^{-5}$. Still, the different ground state crossings (indicated by red dots) between s(inglet), d(oublet) and q(uartet) states lie precisely on top of the peaks of $D_{\mathrm{max}}$. 


In Figs. \ref{fig:four-site-next-nearest-QTTD}(b)-(c), different $\mu/U$-slices are shown for a lower QTT cutoff of $\epsilon=10^{-8}$, demonstrating how well the proposed QTTD  works by producing stable \Dmax{} peaks for different QTT cutoffs ($\epsilon=10^{-5}$ in (a) vs. $\epsilon=10^{-8}$ in (b)-(c)). This shows that time-scale entanglement diagnosed by the growth of  \Dmax \ is system-inherent and visible across all studied correlators.



\begin{figure}
    \centering 
    \includegraphics[width=\linewidth]{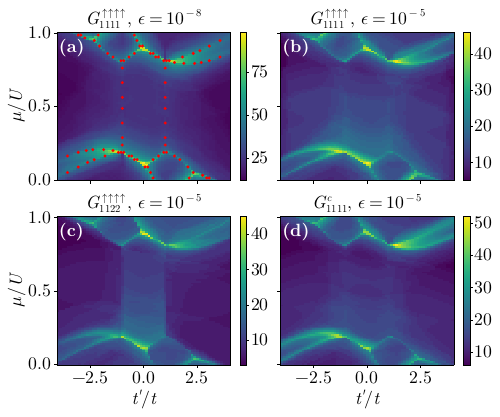}
    \caption{Four-site Hubbard ring with next-nearest neighbor hopping: \Dmax{} of the QTT of various correlators computed with a truncated eigenspectrum for various cutoffs $\epsilon$ at $\beta=50$. The red dots indicate QPTs and lie precisely on top of the maxima of the QTT bond dimensions for all correlators considered.}
    \label{fig:four-site-next-nearest-correlators-QTTD}
\end{figure}

\begin{figure}
    \centering 
    \includegraphics[width=\linewidth]{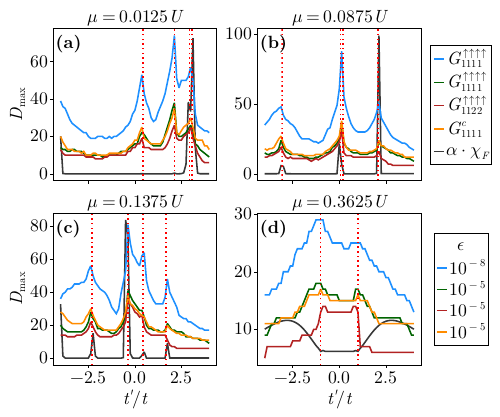}
    \caption{Four-site Hubbard ring with next-nearest neighbor hopping: $\mu$ slices of the \Dmax of the QTT of various correlators computed with a truncated eigenspectrum for various cutoffs $\epsilon$ at $\beta=50$ in comparison to the rescaled value of the fidelity susceptibility $\chi_F$. In (a), (d) the value of $\chi_F$ is rescaled by $\alpha=3\cdot 10^{-1}, 10^2$, while in (b) and (c) $\alpha=1$. The red dashed lines indicate QPTs and lie precisely on top of the sharp peaks of the QTT bond dimensions for all correlators considered.}
    \label{fig:four-site-next-nearest-mu-slices-QTTD}
\end{figure}

Let us now extend these results. In Figs.~\ref{fig:four-site-next-nearest-correlators-QTTD} and~\ref{fig:four-site-next-nearest-mu-slices-QTTD}, we provide results of the QTT bond dimension of various correlators with different SVD cutoffs emphasizing the discussed flexibility provided in the QTTD approach. In Fig.~\ref{fig:four-site-next-nearest-correlators-QTTD}, it can be seen that in the entire parameter space the different correlators
\begin{subequations}
    \begin{align}
    G_{1122}^{\uparrow \uparrow \uparrow \uparrow}(\tau_1, \tau_2, \tau_3) &= - \langle \mathcal{T} \hat c_{1,\uparrow} (\tau_1) \hat c_{1,\uparrow}^{\dagger} (\tau_2) \hat c_{2,\uparrow} (\tau_3) \hat c_{2,\uparrow}^{\dagger} \rangle \\
    G_{1111}^{c}(\tau_1, \tau_2, \tau_3) &= G_{1111}^{\uparrow \uparrow \uparrow \uparrow}(\tau_1, \tau_2, \tau_3) + G_{1111}^{\uparrow \uparrow \downarrow \downarrow}(\tau_1, \tau_2, \tau_3) \\
    G_{1111}^{\uparrow \uparrow \downarrow \downarrow}(\tau_1, \tau_2, \tau_3) &= - \langle \mathcal{T} \hat c_{1,\uparrow} (\tau_1) \hat c_{1,\uparrow}^{\dagger} (\tau_2) \hat c_{1,\downarrow} (\tau_3) \hat c_{1,\downarrow}^{\dagger} \rangle
\end{align}
\end{subequations}
show the same universal peaks. These peaks are stable for different QTT cutoffs (see (a) $\epsilon=10^{-8}$ and (b) $\epsilon=10^{-5}$) identifying physical QPTs. 

This is even better visible in Fig.~\ref{fig:four-site-next-nearest-mu-slices-QTTD}. Here, various $\mu$ slices are presented for the same correlator showing that the sharp QTT bond dimension peaks are precisely located at the QPTs. Even the singlet-singlet crossings for intermediate values of $\mu/U$ become better visible when looking at the $\mu$ slices (Fig.~\ref{fig:four-site-next-nearest-mu-slices-QTTD} (d)). 

\begin{figure}
    \centering 
    \includegraphics[width=\linewidth]{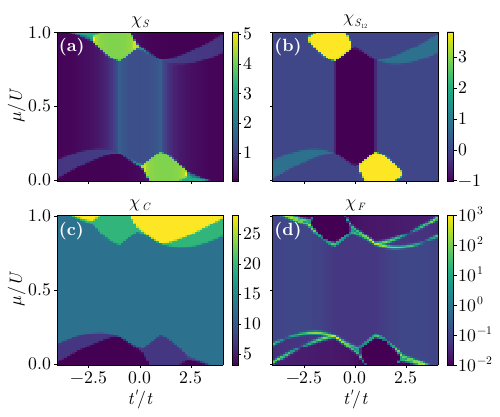}
    \caption{The local spin susceptibility $\chi_s$ (a), the non-local spin susceptibility $\chi_{s_{12}}$ (b), the charge susceptibility $\chi_c$ (c) and the fidelity susceptibility (d) in the case of the four-site Hubbard ring with next-nearest neighbor hopping are presented.}
    \label{fig:four-site-next-nearest-susceptibilities-QTTD}
\end{figure}

It is interesting to compare this result with Fig.~\ref{fig:four-site-next-nearest-susceptibilities-QTTD} (c), where the value of the charge susceptibility $\chi_c$ is shown. Here, the singlet-singlet crossings at intermediate $\mu/U$ values are not visible. However, in Fig.~\ref{fig:four-site-next-nearest-mu-slices-QTTD} (d), these ground state crossings can also be identified in the QTT bond dimension of the ``four-point charge correlator'' $G_{1111}^c$. This means that although this QPT cannot be identified in the physical $\chi_c$, it can be diagnosed with the QTT bond dimension of the ``building block'' of $\chi_c$. This can again be understood from the perspective of the time-scale entanglement inherent in the physical system as a fundamental of QTTD. Moreover, this means that in the case of data of correlators that might not be sufficiently precise in order to draw direct conclusions on possible phase transitions, applying QTTD to these correlators can still lead to the identification of phase transitions and crossovers. Additionally, when investigating the fidelity susceptibility in Fig.~\ref{fig:four-site-next-nearest-mu-slices-QTTD} (d), it can be seen that $\chi_F$ does not detect the discussed singlet-singlet ground state crossings, as no maximum in $\chi_F$ is located at these points. Moreover, let us emphasize that in this panel, the very small value of $\chi_F$ is rescaled by $\alpha = 10^2$ to make it visible in the figure. Hence, in this case QTTD correctly diagnoses these crossings, while $\chi_F$ not only fails to identify these crossings, but also has broader maxima located away from these points that could lead to incorrect conclusions. A second advantage of QTTD over using the fidelity susceptibility becomes clear in Fig.~\ref{fig:four-site-next-nearest-mu-slices-QTTD} (a). In QTTD, the bond dimension maxima are clear and precisely located at the corresponding ground state crossings. When considering the fidelity susceptibility, no peaks are visible at the first and second crossings. However, when investigating the numerical data, maxima can be found at these positions, but the numerical values are suppressed by $\mathcal{O}(10^2-10^3)$ in comparison to the large peak at the third/fourth crossing. While with QTTD all crossings were easily identified due to bond dimension peaks of the same order, the peaks in $\chi_F$ cannot only be easily overlooked, but even if discovered, wrongly dismissed as indicators of ground state crossings due to the different orders of the values of $\chi_F$.

\begin{figure}
    \centering 
    \includegraphics[width=\linewidth]{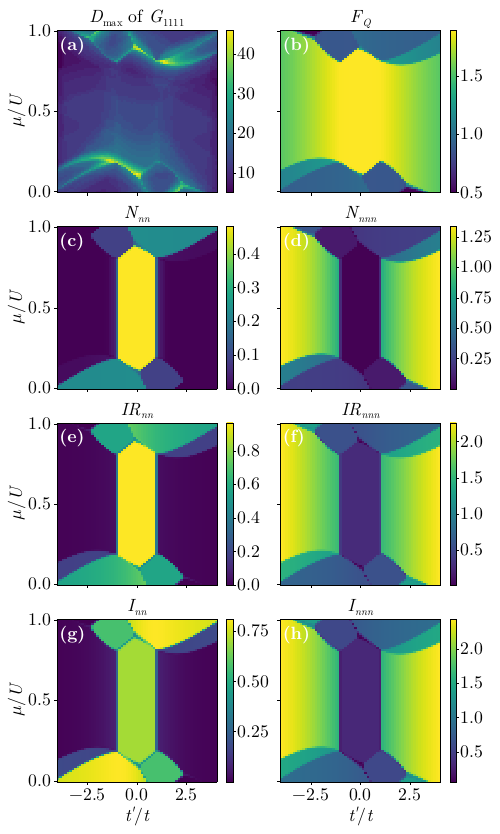}
    \caption{Various entanglement measures in comparison to the \Dmax{}  of the QTT of $G_{1111}^{\uparrow \uparrow \uparrow \uparrow}(\tau_1, \tau_2, \tau_3)$ at $\epsilon=10^{-14}$ at $\beta=50$ for the four-site Hubbard ring with next-nearest neighbor hopping. (b) presents the quantum Fisher information $F_Q$, while (c), (e) and (g) show the nearest-neighbor negativity ($N_{nn}$), Renyi mutual information ($IR_{nn}$) and mutual information ($I_{nn}$). (d), (f) and (h) present the corresponding next-nearest neighbor measures.}
    \label{fig:four-site-next-nearest-entanglement-QTTD}
\end{figure}

Apart from the the susceptibilities shown in Fig.~\ref{fig:four-site-next-nearest-susceptibilities-QTTD} with the non-local susceptibility
\begin{align}
    \chi_{s_{12}} = \int_0^{\beta}d \tau \langle \hat S_{z_1}(\tau) \hat S_{z_2}(0) \rangle,
\end{align}
where $\hat S_{z_i}=(\hat n_{i\uparrow} - \hat n_{i\downarrow})/2$, we also present different entanglement measures in comparison to the \Dmax{}  of $G_{1111}^{\uparrow \uparrow \uparrow \uparrow}$ in Fig.~\ref{fig:four-site-next-nearest-entanglement-QTTD}. Here, we are able to match the maximum in the bond dimension to  jumps and rapid changes in entanglement due to phase transitions, as already seen and discussed in the previous sections.


\section{Mott transition \label{sec:Mott}}
\begin{figure}
    \centering
    \includegraphics[width=1.0\linewidth]{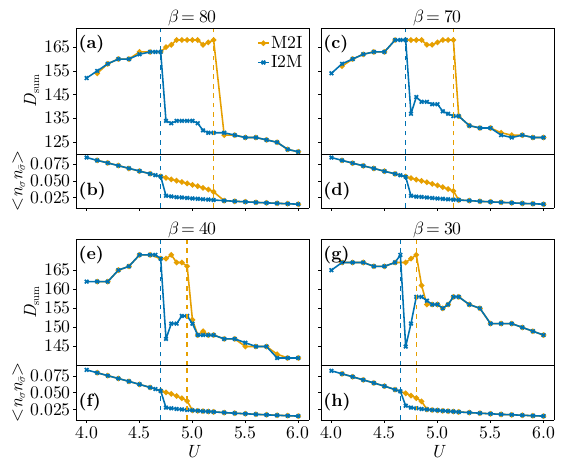}
    \caption{DMFT solution of Hubbard model on Bethe lattice: \Dsum{} of the QTT of $G(i\nu)G(i\nu+i\omega)$ shown for various temperatures as a function of $U$ in comparison to the double occupancy $\langle n_\sigma n_{\bar \sigma} \rangle$ obtained for $R=11,\epsilon=10^{-4}$. Coexistence region is clearly visible form difference of solution initialized from metallic (M2I) and insulating (I2M) regime.}
    \label{fig:Bethe-temperature}
\end{figure}

Here, we extend the analysis of the DMFT results of the Hubbard model on the Bethe lattice and of the DFT+DMFT results of 3D $\mathrm{NdNiO}_2$ provided in the main text. Fig.~\ref{fig:Bethe-temperature} compares the QTT summed bond dimension of the DMFT solution initialized from the metallic (M2I) and insulating (I2M) regime for various temperatures in comparison to the double occupancy. It can be seen that time-scale entanglement increases when approaching the coexistence region.

\begin{figure}
    \centering
    \includegraphics[width=1.0\linewidth]{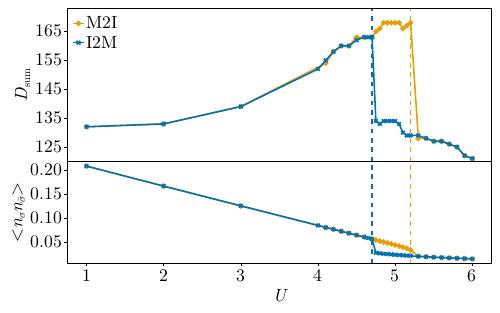}
    \caption{DMFT solution of Hubbard model on Bethe lattice: \Dsum{} of the QTT of $G(i\nu)G(i\nu+i\omega)$ shown for $\beta=90$ as a function of $U$ in comparison to the double occupancy $\langle n_\sigma n_{\bar \sigma} \rangle$ obtained for $R=11,\epsilon=10^{-4}$. The $U$ range is extended until $U=1$ to emphasize the enhancement of time-scale entanglement at the first-order transition. Coexistence region is clearly visible form difference of solution initialized from metallic (M2I) and insulating (I2M) regime.}
    \label{fig:Bethe-beta-90}
\end{figure}

In Fig.~\ref{fig:Bethe-beta-90}, we extend the result at $\beta=90$ already shown in the main text (Fig. 2(a)) to lower values of $U$. Then the clear peak in time-scale entanglement at the the first-order phase transition becomes better visible. This emphasizes that time-scale entanglement is enhanced both on the metallic and insulating side of the transition.

\begin{figure}
    \centering
    \includegraphics[width=1.0\linewidth]{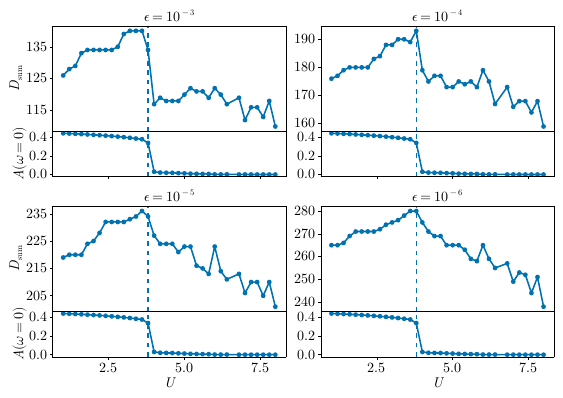}
    \caption{DFT+DMFT solution of 3d $\mathrm{NdNiO}_2$: \Dsum{} of the QTT of the $G(i\nu)G(i\nu+i\omega)$ shown for various QTT cutoffs $\epsilon$ as a function of $U$ in comparison to the the spectral weight $A(\omega)$ at $\omega=0$ obtained for $R=11,\beta=38(\mathrm{eV})$.}
    \label{fig:NNO-cutoff}
\end{figure}

Last, we show in Fig.~\ref{fig:NNO-cutoff} that varying the QTT cutoff $\epsilon$ does not change the qualitative behavior of the sum of bond dimensions across the crossover from metallic to insulating solution showing the enhancement of entanglement between time scales for $\mathrm{NdNiO}_2$.

\section{Single-impurity Anderson model \label{sec:siam}}
\subsection{Discretization and fitting of the bath}
The single-impurity Anderson model (SIAM) is coupled to a bath with a continuous energy spectrum. To perform numerical diagonalization, it is necessary to discretize the bath levels. Here, we describe the details of the discretization procedure.

The Hamiltonian before discretization is given by
\begin{align}
\hat{H} =& (\varepsilon_0 - \mu)(\hat{n}_\uparrow + \hat{n}_\downarrow) + U\hat{n}_\uparrow \hat{n}_\downarrow\nonumber\\
&+ \sum_{\bm{k}\sigma}\varepsilon_{\bm{k}\sigma}\hat{c}_{\bm{k}\sigma}^\dag \hat{c}_{\bm{k}\sigma}
+ \sum_{\bm{k}\sigma}(V_{\bm{k}\sigma}\hat{d}^\dag_{\sigma}\hat{c}_{\bm{k}\sigma} + \mathrm{h.c.}).
\end{align}

Here, the hybridization function is defined using the bath density of states $\rho(\varepsilon)$ as 
\begin{align}
\Delta(\mathrm{i}\omega) = \sum_{\bm{k}}\dfrac{|V_{\bm{k}\sigma}|^2}{\mathrm{i}\omega - \varepsilon_{\bm{k}}} = \int_{-D}^{D}d\varepsilon \dfrac{|V|^2\rho(\varepsilon)}{\mathrm{i}\omega - \varepsilon}.
\end{align}
The density of states of the bath is given by $\rho(\varepsilon) = \frac{2}{\pi D}\sqrt{1 - (\varepsilon/D)^2}$ for the semicircular case, and by $\rho(\varepsilon) = \frac{1}{2D} \theta(\varepsilon + D)\left[1 - \theta(\varepsilon - D)\right]$ for the rectangular case. Here, $\theta(\varepsilon)$ denotes the Heaviside step function, and $D$ is half the bandwidth.
Although $\bm{k}$ is a continuous variable, we discretize the bath into three levels. Let $E_\ell$ and $V_\ell~(\ell=1,2,3)$ denote the discretized bath energies and hybridization strengths, respectively. The Hamiltonian after discretization becomes
\begin{align}
\hat{H} =& (\varepsilon_0 - \mu)(\hat{n}_\uparrow + \hat{n}_\downarrow) + U\hat{n}_\uparrow \hat{n}_\downarrow \nonumber\\
&+ \sum_{\ell=1}^{3}E_{\ell}\hat{c}_{\ell}^\dag \hat{c}_{\ell} + \sum_{\ell=1}^{3}(V_{\ell}\hat{d}^\dag \hat{c}_{\ell} + \mathrm{h.c.}).
\end{align}
The hybridization function after discretization is as follows;
\begin{align}
    \tilde \Delta(\mathrm{i}\omega) = \sum_{\ell=1}^{3}\dfrac{|V_\ell|^2}{\mathrm{i}\omega - E_\ell}.
\end{align}
The bath parameters $V_\ell$ and $E_\ell$ are determined so that the hybridization function before and after discretization matches by minimizing $\|\Delta(\mathrm{i}\omega) - \Tilde{\Delta}(\mathrm{i}\omega)\|^2$.


\subsection{Additional results}

\begin{figure}
    \centering 
    \includegraphics[width=\linewidth]{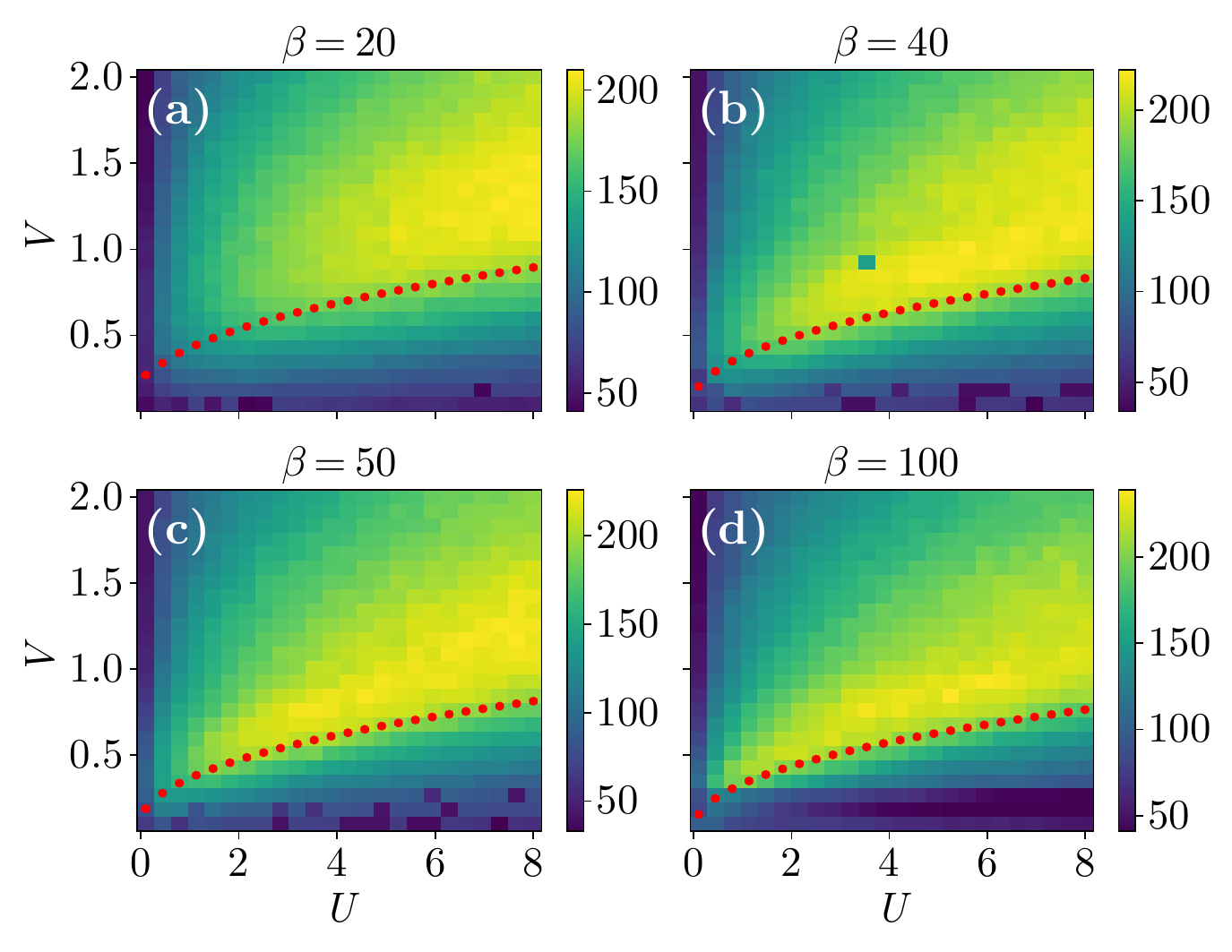}
    \caption{$D_{\mathrm{max}}$ of the QTT for two-particle Green's function $G^{\uparrow\uparrow\uparrow\uparrow}_{\mathrm{imp}}(\tau_1, \tau_2, \tau_3)$ in case of the semicircular density of states. The TCI tolerance is set to $\mathrm{tol}=10^{-5}$ in all temperatures. The red dots indicate the Kondo temperature.}
    \label{fig:SIAM_QTTD_beta20_40_50_100}
\end{figure}

\begin{figure}
    \centering
    \includegraphics[width=\linewidth]{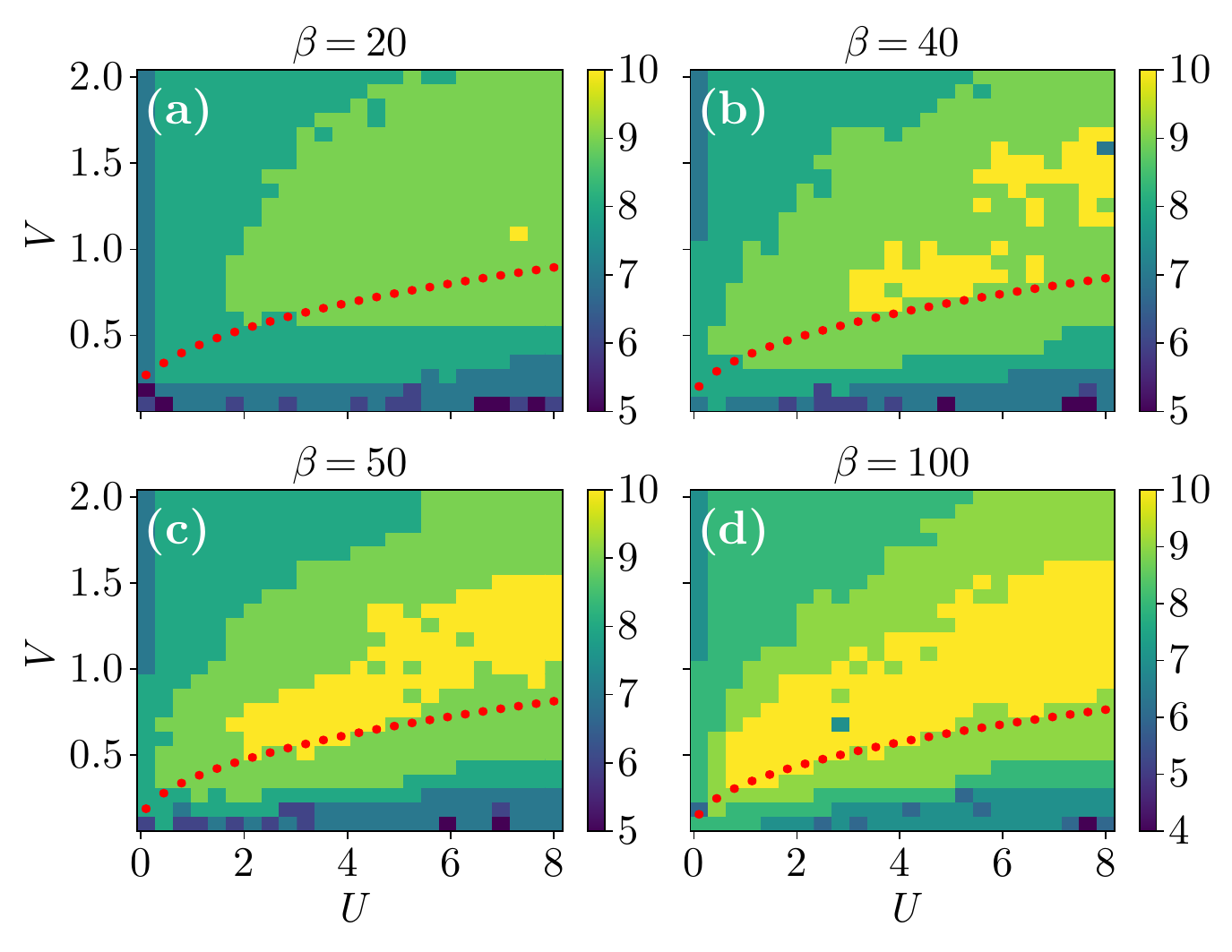}
    \caption{$D_{\mathrm{max}}$ of the QTT for one-particle Green's function $G^{\uparrow\uparrow}_{\mathrm{imp}}(\tau)$ in case of the semicircular density of states. The TCI tolerance is set to $\mathrm{tol}=10^{-10}$ in all temperatures. The red dots indicate the Kondo temperature.}
    \label{fig:SIAM_QTTD_2pcorr_beta20_40_50_100}
\end{figure}

\begin{figure}
    \centering
    \includegraphics[width=\linewidth]{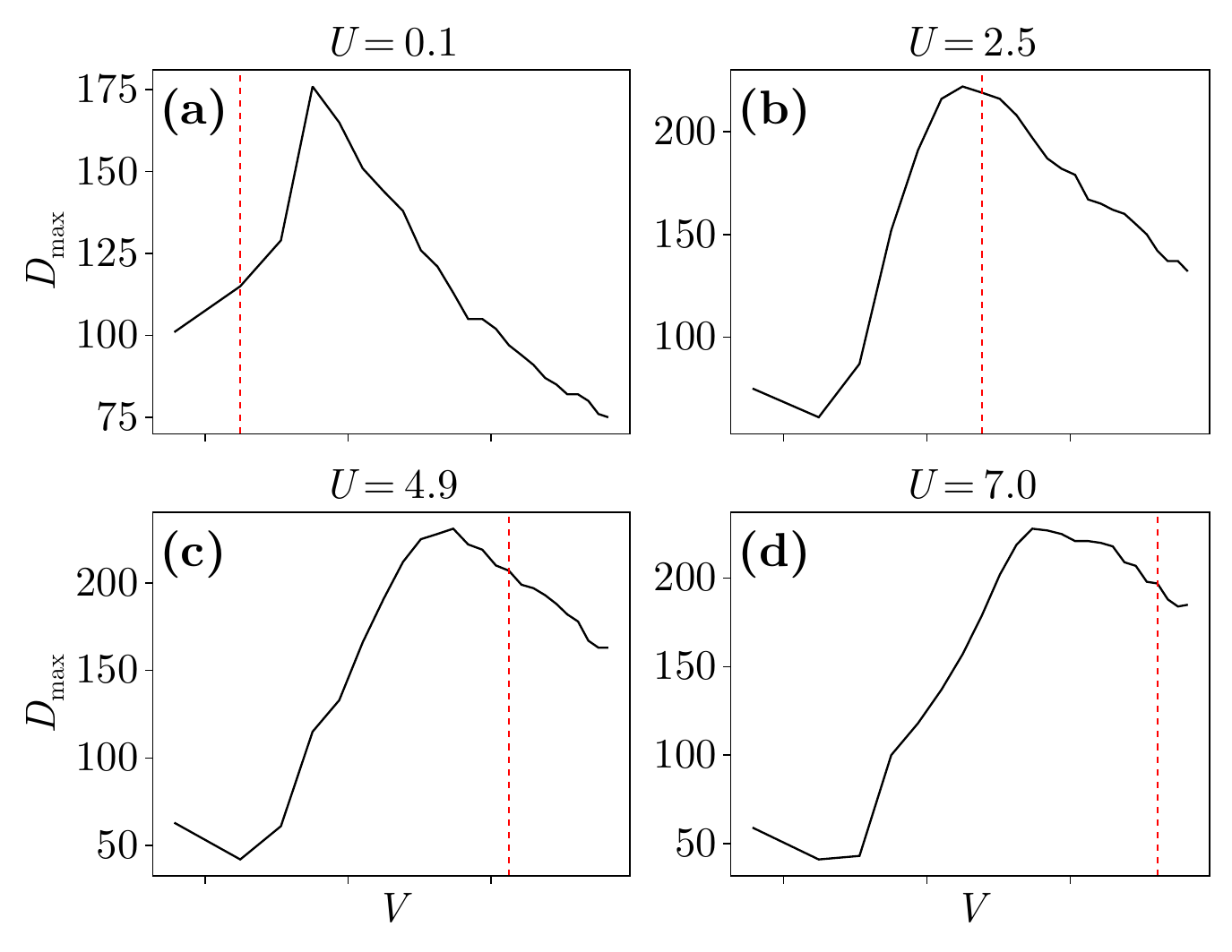}
    \caption{
    U slice of the $D_{\mathrm{max}}$ of the QTT for the two-particle Green's function $G^{\uparrow\uparrow\uparrow\uparrow}_{\mathrm{imp}}(\tau_1, \tau_2, \tau_3)$ in case of the semicircular density of states. The TCI tolerance is set to $\epsilon=10^{-5}$, and the temperature is $\beta=100$. The red dots indicate the Kondo temperature.}
    \label{fig:SIAM_QTTD_4pcorr_Ucut_beta20_40_50_100}
\end{figure}

\begin{figure}
    \centering
    \includegraphics[width=\linewidth]{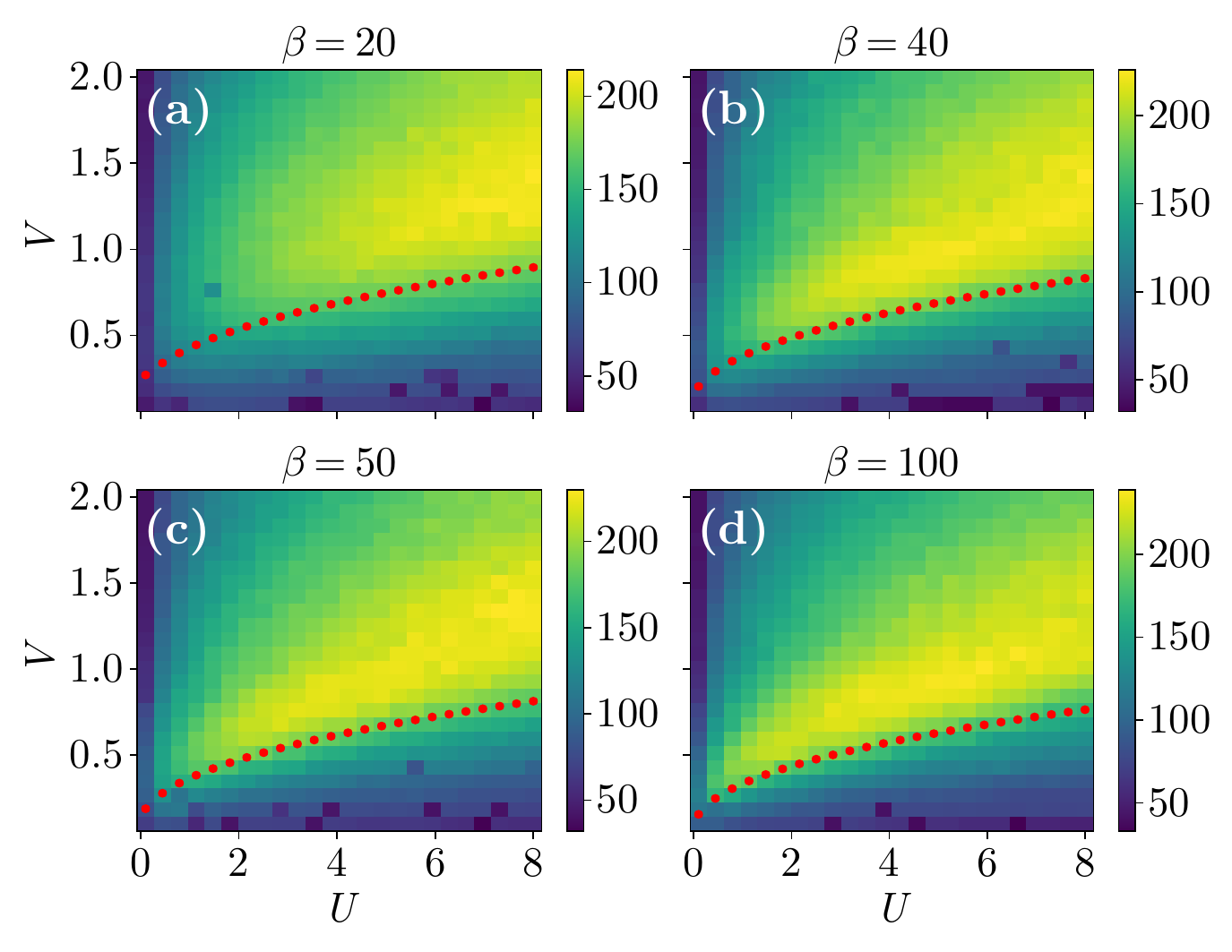}
    \caption{$D_{\mathrm{max}}$ of the QTT for two-particle Green's function $G^{\uparrow\uparrow\uparrow\uparrow}_{\mathrm{imp}}(\tau_1, \tau_2, \tau_3)$ in case of the rectangular density of states. The TCI tolerance is set to $\mathrm{tol}=10^{-5}$ in all temperatures. The red dots indicate the Kondo temperature.}
    \label{fig:SIAM_QTTD_flatDOS_4pcorr_beta20_40_50_100}
\end{figure}

Here we present additional results for the one- and two-particle correlation functions of the SIAM.
Figures~\ref{fig:SIAM_QTTD_beta20_40_50_100}, ~\ref{fig:SIAM_QTTD_2pcorr_beta20_40_50_100} and~\ref{fig:SIAM_QTTD_4pcorr_Ucut_beta20_40_50_100} show the correlation functions for a semicircular density of states.
Figure~\ref{fig:SIAM_QTTD_beta20_40_50_100} illustrates the dependence of the maximum bond dimension of the correlation functions on model parameters at four different temperatures.
At all the temperatures shown, the maximum bond dimension captures the crossover from the local moment regime to the Kondo regime.
Figure~\ref{fig:SIAM_QTTD_4pcorr_Ucut_beta20_40_50_100} shows a slice of the data at $\beta = 100$ as a function of interaction strength U. As can be seen, the peak in the maximum bond dimension is broad. Although the peak cannot be resolved precisely due to the limited bond dimension, its location is well captured.
Figure~\ref{fig:SIAM_QTTD_2pcorr_beta20_40_50_100} shows the maximum bond dimension of the one-particle correlation function. While the maximum bond dimension is significantly smaller than that for the two-particle case, it shows qualitatively similar behavior in the plots.

Finally, Fig.~\ref{fig:SIAM_QTTD_flatDOS_4pcorr_beta20_40_50_100} presents the maximum bond dimension of the two-particle correlation function in the case of a rectangular DOS. The results show behavior similar to that of the semicircular density of states, which is expected, as the dominant contributions come from electrons near the Fermi surface, and are thus largely independent of the detailed shape of the density of states.

\section{Transverse-field Ising model \label{sec:tfim}}
\begin{figure}
    \centering
    \includegraphics[width=\linewidth]{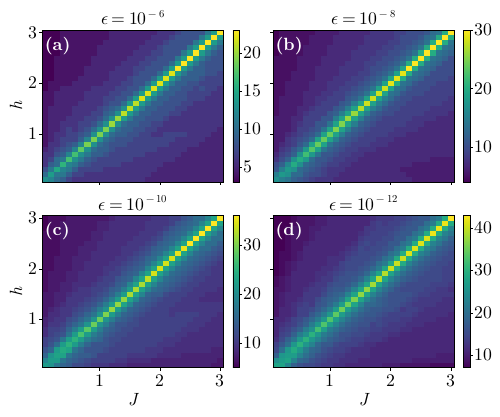}
    \caption{\Dmax{} of $G(k,i\nu)$ of the TFIM for $R=12$ at $\beta=10^4$ shown for various QTT cutoffs $\epsilon$.}
    \label{fig:tfim-G-beta-e4-maxdim}
\end{figure}

\begin{figure}
    \centering
    \includegraphics[width=\linewidth]{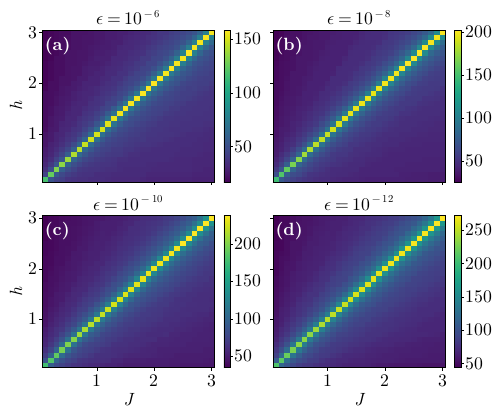}
    \caption{\Dsum{} of $G(k,i\nu)$ of the TFIM for $R=12$ at $\beta=10^4$ shown for various QTT cutoffs $\epsilon$.}
    \label{fig:tfim-G-beta-e4-sumdim}
\end{figure}

In this section, we provide additional results supporting the conclusions reached in the main text in the case of the transverse-field Ising model. First, in Figs.~\ref{fig:tfim-G-beta-e4-maxdim} and~\ref{fig:tfim-G-beta-e4-sumdim}, we show the maximum bond dimension \Dmax{} and the sum of bond dimensions \Dsum{} of the QTT compressed normal Green's function $G(k,i\nu)$ for various QTT cutoffs $\epsilon$. Independent of the cutoff $\epsilon$, there is a peak in \Dmax{} and even sharper in \Dsum{} at the quantum phase transition $h=J$.

\begin{figure}
    \centering
    \includegraphics[width=\linewidth]{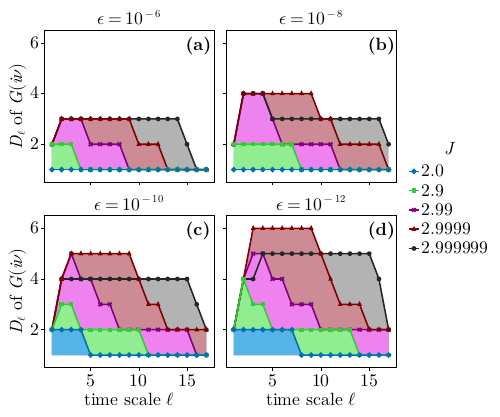}
    \caption{Bond dimension profile $D_\ell$ of local Green's function $G(i \nu)$ at $\beta=10^7, R=18$ near criticality ($h=3$) for various QTT cutoffs $\epsilon$. Criticality and scale invariance lead to uniform entanglement across logarithmic time scales.}
    \label{fig:tfim-G-local-supp}
\end{figure}

Similarly, the bond dimension structure of the QTT compressed local Green's function $G(i\nu)$ remains qualitatively independent of the QTT cutoff $\epsilon$ when approaching the quantum phase transition. This becomes visible in Fig.~\ref{fig:tfim-G-local-supp}, where near the quantum critical point all scales contribute equally.

\section{Data availability \label{sec:data}}
A data set containing all numerical data and plot scripts used to generate the figures of this publication is publicly
available at XXX.

\section{Computational details \label{sec:computation}}
All QTT calculations were performed in \verb|Julia|. The calculation of the various correlators and susceptibilities was performed with exact diagonalization. The compression into QTTs was conducted using the \verb|ITensors|, \verb|QuanticsGrids| and \verb|TensorCrossInterpolation| libraries. In the SVD-based compression the cutoff is based on the squared Frobenius norm, while the tolerance in the TCI-based QTT compression is based on the maximum norm. DFT+DMFT computations were performed with Wien2K~\cite{wien2k}, Wien2wannier~\cite{wien2wannier}, Wannier90~\cite{wannier90} and w2dynamics~\cite{w2dynamics}. Analytic continuation was performed using the ana-cont~\cite{ana_cont} package.

\bibliography{Bibliography_QTTdiagnostics}